\documentclass[12pt]{article}

\usepackage{mysettings}

\title{Probabilistic Count Matrix Factorization for Single Cell Expression Data Analysis}
\author{G.~Durif\,$^{\text{1,2,3,}*}$, L.~Modolo\,$^{\text{1,4,5}}$, J. E. Mold\,$^{\text{5}}$, S.~Lambert-Lacroix\,$^{\text{6}}$ and F.~Picard\,$^{\text{1}}$}
\date{\today}

\begin{document}
\setlength{\parindent}{0pt}
\maketitle

$^{\text{\sf 1}}$Univ Lyon, Universit\'e Lyon 1, CNRS, LBBE UMR 5558, Villeurbanne, France\\
$^{\text{\sf 2}}$Univ Grenoble Alpes, Inria, CNRS, Grenoble INP, LJK UMR 5224, Grenoble, France, \\
$^{\text{\sf 3}}$Universit\'e de Montpellier, CNRS, IMAG UMR 5149, Montpellier, France, \\
$^{\text{\sf 4}}$Univ Lyon, ENS Lyon, Universit\'e Lyon 1, CNRS, LBMC UMR 5239, Lyon, France,\\
$^{\text{\sf 5}}$Department of Cell and Molecular Biology, Karolinska Institutet, Stockholm, Sweden, \\
$^{\text{\sf 6}}$Univ Grenoble Alpes, CNRS, TIMC-IMAG UMR 5525, Grenoble, France.\\
$^{*}$Corresponding author: \url{ghislain.durif@umontpellier.fr}

\begin{abstract}
\textbf{Motivation:} The development of high throughput single-cell sequencing technologies now allows the investigation of the population diversity of cellular transcriptomes. The expression dynamics (gene-to-gene variability) can be quantified more accurately, thanks to the measurement of lowly-expressed genes. In addition, the cell-to-cell variability is high, with a low proportion of cells expressing the same genes at the same time/level. Those emerging patterns appear to be very challenging from the statistical point of view, especially to represent a summarized view of single-cell expression data. PCA is a most powerful tool for high dimensional data representation, by searching for latent directions catching the most variability in the data. Unfortunately, classical PCA is based on Euclidean distance and projections that poorly work in presence of over-dispersed count data with dropout events like single-cell expression data.\\
\textbf{Results:} We propose a probabilistic Count Matrix Factorization (pCMF) approach for single-cell expression data analysis, that relies on a sparse Gamma-Poisson factor model. This hierarchical model is inferred using a variational EM algorithm. It is able to jointly build a low dimensional representation of cells and genes. We show how this probabilistic framework induces a geometry that is suitable for single-cell data visualization, and produces a compression of the data that is very powerful for clustering purposes. Our method is competed against other standard representation methods like t-SNE, and we illustrate its performance for the representation of single-cell expression (scRNA-seq) data.\\
\textbf{Availability:} Our work is implemented in the pCMF R-package\footnote{\url{https://github.com/gdurif/pCMF}}.
\end{abstract}

\setlength{\parindent}{0pt}
\setlength{\parskip}{\baselineskip}

\newpage
\section{Introduction}

The combination of massive parallel sequencing with high-throughput cell biology technologies has given rise to the field of single-cell genomics, which refers to techniques that now provide genome-wide measurements of a cell's molecular profile either based on DNA \citep{ZLC12}, RNA \citep{PBF13}, or chromatin \citep{BWL15,RRS15}. Similar to the paradigm shift of the 90s characterized by the first molecular profiles of tissues \citep{GST99}, it is now possible to characterize molecular heterogeneities at the cellular level \citep{deng2014,saliba2014}. A tissue is now viewed as a population of cells of different types, and many fields have now identified intra-tissue heterogeneities, in T cells \citep{BNC15}, lung cells \citep{TCG14}, or intestine cells \citep{grun2015}. The construction of a comprehensive atlas of human cell types is now within our reach \citep{WRY16}. The characterization of heterogeneities in single-cell expression data thus requires an appropriate statistical model, as the transcripts abundance is quantified for each cell using read counts. Hence, standard methods based on Gaussian assumptions are likely to fail to catch the biological variability of lowly expressed genes, and Poisson or Negative Binomial distributions constitute an appropriate framework \cite[Chap. 6]{chen2016, riggs2017}. Moreover, dropouts, either technical (due to sampling difficulties) or biological (no expression or stochastic transcriptional activity), constitute another major source of variability in scRNA-seq (single-cell RNA-seq) data, which has motivated the development of the so-called Zero-Inflated models \citep{kharchenko2014}.

Principal component analysis (PCA) is one of the most widely used dimension reduction technique, as it allows the quantification and visualization of variability in massive datasets. It consists in approximating the observation matrix $\mb X_{[n \times m]}$ ($n$ cells, $m$ genes), by a factorized matrix of reduced rank, denoted $\mb U\tr{\mb V}$ where $\mb U_{[n\times K]}$ and $\mb V_{[m\times K]}$ represent the latent structure in the observation and variable spaces respectively. This projection onto a lower-dimensional space (of dimension $K$) allows one to catch gene co-expression patterns and clusters of individuals. PCA can be viewed either geometrically or through the light of a statistical model \citep{landgraf2015}. Standard PCA is based on the $\ell_2$ distance as a metric and is implicitly based on a Gaussian distribution \citep{eckart1936}. Model-based PCA offers the unique advantage to be adapted to the data distribution. It consists in specifying the distribution of the data $\mb X_{[n \times m]}$ through a statistical model, and to factorize $\mathbb{E}(\mb X)$ instead of $\mb X$. In this context the $\ell_2$ metric is replaced by the Bregman divergence which is adapted to maximum likelihood inference \citep{collins2001}. A probabilistic version of the Gaussian PCA was proposed by \cite{pierson2015} in the context of single cell data analysis, with the modeling of zero inflation (the ZIFA method). \revise{However, scRNA-seq}{ScRNA-seq} data may be better analyzed by methods dedicated to count data such as the Non-negative Matrix Factorization (NMF) introduced in a Poisson-based framework by \cite{lee1999} or the Gamma-Poisson factor model \citep{cemgil2009,fevotte2009,landgraf2015}. \revise{However, none}{None} of the currently available dimension reduction methods fully model single-cell expression data, characterized by over-dispersed zero inflated counts \citep{kharchenko2014,zappia2017}.

Our method is based on a probabilistic count matrix factorization (pCMF). We propose a dimension reduction method that is dedicated to over-dispersed counts with dropouts, in high dimension. In particular, gene expression can be normalized but does not require to be transformed (log, Anscombe) in our framework. Our factor model takes advantage of the Poisson Gamma representation to model counts from scRNA-seq data \citep{zappia2017}. In particular, we use Gamma priors on the distribution of principal components. We model dropouts with a Zero-Inflated Poisson distribution \citep{simchowitz2013}, and we introduce sparsity in the model thanks to a spike-and-slab approach \citep{malsiner-walli2011} that is based on a two component sparsity-inducing prior on loadings \citep{titsias2011}. \revise{}{We propose a heuristic to initialize the sparsity layer based on the variance of the recorded variables, acting as an integrated gene filtering step, which is an important issue in scRNA-seq data analysis \citep{soneson2018}.} The model is inferred using a variational EM algorithm that scales favorably to data dimension compared with  Markov Chain Monte-Carlo (MCMC) methods \citep{hoffman2013,blei2017}. Then we propose a new criterion to assess the quality of fit of the model to the data, as a percentage of explained deviance, following a strategy that is standard in the Generalized Linear Models framework. Moreover, we show that our criterion \revise{resumes}{corresponds} to the percentage of explained variance in the PCA case, which makes it suitable to compare geometric and probabilistic methods. 

We show the performance of pCMF on simulated and experimental datasets, in terms of visualization and quality of fit. Moreover, we show the benefits of using pCMF as a preliminary dimension reduction step before clustering or before the popular t-SNE approach \citep{vandermaaten2008,ADT13}. Experimental published data are used to show the capacity of pCMF to provide a better representation of the heterogeneities within scRNA-Seq datasets, which appears to be extremely helpful to characterize cell types. Finally, our approach also provides a lower space representation for genes (and not only for cells), contrary to t-SNE. pCMF is implemented in the form of a R package available at \url{https://github.com/gdurif/pCMF}.

\section{Count Matrix Factorization for zero-inflated over-dispersed data}\label{sec:method}

\paragraph{The Poisson factor model.} Our data consist of a matrix of counts (potentially normalized but not transformed), denoted by $\mb X \in \mathbb{N}^{n\times m}$, that we want to decompose onto a subspace of dimension $K$ (being fixed). In a first step we suppose that the data follow a multivariate Poisson distribution of intensity $\mb \Lambda$. Following the standard Poisson Non-Negative Matrix Factorization \citep[Poisson NMF,][]{lee1999}, we approximate this intensity such that 
\begin{equation}
\mb X \sim \Poi \big(\mb \Lambda\big), \quad \mb \Lambda \simeq \mb U \mb V^T,
\label{eq:data:poisson}
\end{equation}
with factor $\mb U \in \mathbb{R}^{+, n \times K}$ the coordinates of the $n$ observations (cells) in the subspace of dimension $K$, and loadings $\mb V \in \mathbb{R}^{+,m \times K}$ the contributions of the $m$ variables (genes).

\paragraph{Modeling over-dispersion.} We account for over-dispersion by using the the Negative-Binomial distribution  \citep{anders2010}, through a hierarchical Gamma-Poisson representation (GaP) \cite{cemgil2009}. In our factor model $\mb U$ and $\mb V$ are modeled as independent random latent variables with Gamma distributions such that
\begin{equation}
\begin{aligned}
& U_{ik} \sim \Gam(\alpha_{k,1},\alpha_{k,2}) \ \text{for any $(i,k)\in[1:n] \times [1:K]$} \, , \\
& V_{jk} \sim \Gam(\beta_{k,1},\beta_{k,2}) \ \text{for any $(j,k)\in[1:m] \times [1:K]$} \, .\\
\end{aligned}
\label{eq:factor:gamma}
\end{equation}
In practice, some third-party latent variables are introduced for the derivation of our inference algorithm \citep{cemgil2009, zhou2012}. We consider latent variables $\mb Z = [Z_{ijk}]\in\RR^{n\times m \times K}$, defined such that $X_{ij} = \msum_k\, Z_{ijk}$. These new indicator variables quantify the contribution of factor $k$ to the data. Here $Z_{ijk}$ are assumed to be conditionally independent and to follow a conditional Poisson distribution, i.e. $Z_{ijk}\,\vert\,U_{ik},V_{jk} \sim \Poi(U_{ik}\,V_{jk})$. Thus, the conditional distribution of $X_{ij}$ remains $\Poi(\msum_k\, U_{ik}\,V_{jk})$ thanks to the additive property of the Poisson distribution.

\paragraph{Dropout modeling with a zero-inflated (ZI) model.} To model zero-inflation, i.e. random null observations called dropout events, we introduce a dropout indicator variable $D_{ij} \in\{0,1\}$ for $i=1,\dots,n$ and $j=1,\dots,p$ \citep[c.f.][]{simchowitz2013}. In this context, each $D_{ij}=0$ if gene $j$ has been subject to a dropout event in cell $i$, with $D_{ij} \sim \mathcal{B}(\pi_j^D)$.
We consider gene-specific dropout rates, $\pi_j^D$, following recommendations of the literature \citep{pierson2015}. Thus, to include zero-inflation in the probabilistic factor model, we consider that
\[
X_{ij}\,|\,\mb U_{i}, \mb V_{j}, \mb D \sim (1-D_{ij})\times \delta_0+D_{ij} \times \Poi\big(\sum_{k}\,U_{ik}\,V_{jk}\big)\, ,
\]
\revise{}{where $\delta_0$ is the Dirac mass at 0, i.e. $\delta_0(X_{ij}) = 1$ if $X_{ij}=0$ and 0 otherwise.} The dropout indicators $D_{ij}$ are assumed to be independent from the factors $U_{ik}$ and $V_{jk}$. Then, by integrating $D_{ij}$ out, the probability of observing a zero in the data illustrates the two potential sources of zeros and becomes
\[
\PP\big(X_{ij} = 0\,\vert\,\mb U_i, \mb V_j\,;\,\mbg \pi \big) = (1-\pi_j^D) + \pi_j^D\, \exp\big( - \msum_k U_{ik}\,V_{jk}\big).
\]
\revise{}{Thus, inference will be based on the factors $U_{ik}$ and $V_{jk}$ and on probabilities $\pi_j^D$.}

\paragraph{Probabilistic variable selection.} Finally we suppose that our model is parsimonious. We consider that among all recorded variables, only a proportion carries the signal and the others are noise. To do so, we modify the prior of the loadings variables $V_{jk}$, to consider a sparse model with a two-group sparsity-inducing prior \citep{engelhardt2014}. The model is then enriched by the introduction of a new indicator variable $S_{jk} \sim \mathcal{B}(\pi_{j}^S)$, that equals 1 if gene $j$ contributes to loading $V_{jk}$, and zero otherwise. $\pi_{j}^S$ stands for the prior probability for gene $j$ to contribute to any loading. To define the sparse GaP factor model, we modify the distribution of the loadings latent factor $V_{jk}$, such that  
\[ V_{jk}|S_{jk} \sim (1-S_{jk}) \times \delta_0\, + S_{jk} \times  \,\Gam(\beta_{k,1}, \beta_{k,2})\, .\]
This spike-and-slab formulation \citep{mitchell1988} ensures that $V_{jk}$ is either null (gene $j$ does not contribute to factor $k$), or drawn from the Gamma distribution (when gene $j$ contributes to the factor). The contribution of gene $j$ to the component $k$ is accounted for in the conditional Poisson distribution of $X_{ij}$, with 
\[
\begin{aligned}
X_{ij}\,\vert \mb U_{i}, {\mb V}_{j}', \mb D, \mb S_{j} & \sim && (1-D_{ij})(1-S_{jk})\times \delta_0 \\
&&& + \Poi\big(D_{ij}\msum_k\, \,U_{ik}\,[S_{jk}\,V'_{jk}]\big),
\end{aligned}
\]
where $V_{jk} = S_{jk}\,V'_{jk}$ such that $V'_{jk}\sim \Gam(\beta_{k,1}, \beta_{k,2})$.

\paragraph{Underlying geometry.}\label{supp:sec:geometry} Knowing $\mb U$ and $\mb V$, to quantify the approximation of matrix $\mb X$ by $\mb U\tr{\mb V}$, we consider the Bregman divergence, that can be viewed as a generalization of the Euclidean metric to the exponential family \cite[see][]{collins2001,banerjee2005, chen2008}. In the Poisson model, the Bregman divergence between $\mb X$ and $\mb U\tr{\mb V}$ is defined as \citep{fevotte2009}:
\[
D(\mb X\,\vert\,\mb U\tr{\mb V}) = \sum_{i=1}^n\sum_{j=1}^m x_{ij}\,\log\left(\frac{x_{ij}}{\sum_k U_{ik}V_{jk}}\right) - x_{ij} + \sum_k U_{ik}V_{jk}.
\]
Hence the geometry is induced by an appropriate probabilistic model dedicated to count data. Potential identifiability issues are addressed in Supp. Mat. (\cref{supp:sec:identifiability}).

\revise{}{In the following, we will refer to pCMF for the method implementing the model with dropout but the without sparsity layer, and to sparse pCMF (or spCMF) for the model with dropout and sparsity layers.}

\subsection{Quality of the reconstruction.}\label{sec:method:expdev}

The Bregman divergence between the data matrix $\mb X$ and the reconstructed matrix $\hat{\mb U}\tr{\hat{\mb V}}$ in our GaP factor model is related to the deviance of the Poisson model defined such as \citep{landgraf2015}
\[
\text{Dev}(\mb X, \hat{\mb U}\tr{\hat{\mb V}}) = -2 \times \big(\log p(\mb X\,|\,\mbg\Lambda = \hat{\mb U}\tr{\hat{\mb V}}) - \log p(\mb X\,|\,\mbg\Lambda = \mb X)\big)\, ,
\]
where $\log p(\mb X\,|\,\mbg\Lambda)$ is the Poisson log-likelihood based on the matrix notation~(\ref{eq:data:poisson}). We have  $\text{Dev}(\mb X, \hat{\mb U}\tr{\hat{\mb V}}) \propto D(\mb X\,\vert\,\hat{\mb U}\tr{\hat{\mb V}})$, thus the deviance can be used to quantify the quality of the model.

Regarding PCA, the percentage of explained variance is a natural and unequivocal quantification of the quality of the representation. We introduce a criterion that we call percentage of explained deviance that is a generalization of the percentage of explained variance to our GaP factor model. However, since our models are not nested for increasing $K$, the definition of this criterion appears non trivial. To assess the quality of our model, we propose to define the percentage of explained deviance as:
\begin{equation}
\%\text{dev} = \frac{\log p(\mb X\,\vert\,\mbg\Lambda = \hat{\mb U}\tr{\hat{\mb V}}) - \log p(\mb X\,\vert\,\mbg\Lambda = \II_n \bar{\mb X})}{\log p(\mb X\,\vert\,\mbg\Lambda = \mb X) - \log p(\mb X\,\vert\,\mbg\Lambda = \II_n \bar{\mb X})}
\label{eq:exp_deviance}
\end{equation}
where $\hat{\mb U}\tr{\hat{\mb V}}$ is the predicted reconstructed matrix in our model, $\II_n$ is a column vector filled with 1 and $\bar{\mb X}$ is a row vector of size $m$ storing the column-wise average of $\mb X$. We use two baselines: $(i)$ the log-likelihood of the saturated model, i.e. $\log p(\mb X\,\vert\,\mbg\Lambda = \mb X)$ (as in the deviance), which corresponds to the richest model and $(ii)$ the log-likelihood of the model where each Poisson intensities $\lambda_{ij}$ is estimated by the average of the observations in the column $j$, i.e. $\log p(\mb X\,\vert\,\mbg\Lambda = \II_n \bar{\mb X})$, which is the most simple model that we could use. This formulation ensures that the ratio $\%\text{dev}$ lies in $[0;1]$. An interesting point is that if we assume a Gaussian distribution on the data, the percentage of explained deviance is exactly the percentage of explained variance from PCA (c.f. \cref{supp:sec:exp_deviance}), which makes our criterion suitable for to compare different factor models.

\subsection{Choosing the dimension of the latent space}

As noticed in the previous section, the GaP factor model with an increasing number $K$ of factors are not nested (the model associated to the NMF presents the same properties). Consequently, testing different values of $K$ requires to fit different models (contrary to PCA for instance). We choose the number of factors by fitting a model with a large $K$ and \revise{verify}{verifying} how the matrix $\hat{\mb U}_{1:k}\tr{(\hat{\mb V}_{1:k})}$ reconstructs $\mb X$ depending on $k=1,\hdots,K$ \revise{}{with a rule of thumb based on the ``elbow'' shape of the fitting criterion}. This approach is for instance widely used in PCA by checking the proportion of variability explained by each components, see \citet[p.96]{friguet2010} for a review of the different criteria to choose $K$ in this context. Here we use the deviance, or equivalently the Bregman divergence $k \mapsto D\big(\mb X\,\vert\,\hat{\mb U}_{1:k}\tr{(\hat{\mb V}_{1:k})}\big)$ to find the latent dimension from where adding new factors does not improve $D\big(\mb X\,\vert\,\hat{\mb U}_{1:k}\tr{(\hat{\mb V}_{1:k})}\big)$. This determination is however not always unambiguous and may sometimes lead to some over-fitting, i.e. when considering too much factors. \revise{}{In addition, when focusing on data visualization, we generally set $K=2$.}

\section{Model inference using a variational EM algorithm}

Our goal is to infer the posterior distributions over the factors $\mb U$ and $\mb V$ depending on the data $\mb X$. To avoid using the heavy machinery of MCMC \citep{nathoo2013} to infer the intractable posterior of the latent variables in our model, we use the framework of variational inference \citep{hoffman2013}. In particular, we extend the version of the variational EM algorithm \citep{beal2003} proposed by \cite{dikmen2012} in the context of the standard Gamma-Poisson factor model to our sparse and zero-inflated GaP model. \cref{supp_mat:fig:varinf} in Supp. Mat. gives an overview of the variational framework.

\subsection{Definition of variational distributions}

In the variational framework, the posterior $p(\mb Z, \mb U,\mb V', \mb S,\mb D\,|\,\mb X)$ is approximated by the variational distribution $q(\mb Z, \mb U, \mb V', \mb S, \mb D)$ regarding the Kullback-Leibler divergence \citep{hoffman2013}, that quantifies the divergence between two probability distributions. Since the posterior is not explicit, the inference of $q$ is based on the optimization of the Evidence Lower Bound (ELBO), denoted by $J(q)$ and defined as:
\begin{equation}
	\begin{aligned}
		J(q) = \EE_q[\log p(\mb X, \mb Z, \mb U,\mb V' ,\mb S, \mb D)] -\EE_q[\log q(\mb Z, \mb U, \mb V', \mb S, \mb D)] \, ,
	\end{aligned}
	\label{eq:elbo}
\end{equation}
that is a lower bound on the marginal log-likelihood $\log p(\mb X)$. In addition, maximizing the ELBO $J(q)$ is equivalent to minimizing the KL divergence between $q$ and the posterior distribution of the model \citep{hoffman2013}. To derive the optimization, $q$ is assumed to lie in the mean-field variational family, i.e. $(i)$ to be factorisable with independence between latent variables and between observations and $(ii)$ to follow the conjugacy in the exponential family, i.e. to be in the same exponential family as the full conditional distribution on each latent variables in the model. Thanks to the first assumption, in our model, the variational distribution $q$ is defined as follows:
\begin{equation}
	\begin{aligned}
		& q(\mb Z, \mb U, \mb V', \mb S, \mb D) = \prod_{i=1}^n\prod_{j=1}^m q\big((Z_{ijk})_k\,|\, (r_{ijk})_k\big) \\
		& \times \prod_{i=1}^n\prod_{k=1}^K q(U_{ik}\,|\,\mb a_{ik}) 
		\times \prod_{j=1}^m \prod_{k=1}^K q(V'_{jk}\,|\,\mb b_{jk}) \\
		& \times \prod_{j=1}^m \prod_{k=1}^K q(S_{jk}\,|\, p_{jk}^S)
		\times \prod_{i=1}^n \prod_{j=1}^m q(D_{ij}\,|\, p_{ij}^D)
	\end{aligned}
\end{equation}
where $(r_{ijk})_k$, $\mb a_{ik}$, $\mb b_{jk}$, $p_{jk}^S$ and $p_{ij}^D$ are the parameters of the variational distribution regarding $(Z_{ijk})_k$, $U_{ik}$, $V'_{jk}$, $S_{jk}$, $D_{ij}$ respectively. Then we need to precise the full conditional distributions of the model before defining the variational distributions more precisely.

\subsection{Approximate posteriors}

To approximate the (intractable) posterior distributions, variational distributions are assumed to lie in the same exponential family as the corresponding full conditionals and to be independent such that: 
\[
\mb Z_{ij} \overset{q}\sim \multin\Big( ( r_{ijk})_k\Big) \ \ \ \ \ \ \ \ \
\begin{aligned}
& U_{ik} \overset{q}\sim \Gam(a_{ik,1},a_{ik,2})\\
& V'_{jk} \overset{q}\sim \Gam(b_{jk,1},b_{jk,2})\\
\end{aligned} \ \ \ \ \ \ \ \ \
\begin{aligned}
& S_{jk} \overset{q}\sim \mathcal{B}(p_{jk}^S)\\
& D_{ij} \overset{q}\sim \mathcal{B}( p_{ij}^D),
\end{aligned}
\]
where $\overset{\scriptscriptstyle q}{\sim}$ denotes the variational distribution. The strength of our approach is the resulting explicit approximate distribution on the loadings that induces sparsity:
\[
V_{jk}|S_{jk} \overset{q}\sim (1-S_{jk}) \times \delta_0 + S_{jk} \times \Gam(b_{jk,1},b_{jk,2}),
\]
In the following, the derivation of variational parameters involves the moments and log-moments of the latent variables regarding the variational distribution. Since the distributions q is fully determined, these moments can be directly computed. For the sake of simplicity, we will use notation $\hat{U}_{ik} = \EE_q[U_{ik}]$ and $\hat{\log U}_{ik} = \EE_q[\log U_{ik}]$ (collected in the matrices $\hat{\mb U}$ and $\hat{\log \mb U}$ respectively), with similar notations for other hidden variables of the model ($V'_{jk}$, $D_{ij}$, $S_{jk}$, $Z_{ijk}$).

\subsection{Derivation of variational parameters}\label{sec:meth:var_update}

In order to find a stationary point of the ELBO, $J(q)$ is differentiated regarding each variational parameter separately. The formulation of the ELBO regarding each parameter separately is based on the corresponding full conditional, e.g. $p(U_{ik}\,\vert\,\cond\,)$, $p(V_{jk}\,\vert\,\cond\,)$ and $p \big((Z_{ijk})_k\,\vert\,\cond\,\big)$. The partial formulation are therefore respectively:
\[
\begin{aligned}
& J(q)\big\vert_{\mb a_{ik}} = \EE_q[\log p(U_{ik}\,\vert\,\cond)] - \EE_q[\log q(U_{ik}\,;\,\mb a_{ik})]  + \text{cst}\\
& J(q)\big\vert_{\mb b_{jk}} = \EE_q[\log p(V'_{jk}\,\vert\,\cond)] - \EE_q[\log q(V'_{jk}\,;\,\mb b_{jk})]  + \text{cst}\\
& \begin{aligned}
& J(q)\big\vert_{(r_{ijk})_k} = && \EE_q\big[\log p\big((Z_{ijk})_k\,\vert\,\cond\big)\big] \\ 
&&& - \EE_q\big[\log q\big((Z_{ijk})_k\,;\,(r_{ijk})_k\big)\big] + \text{cst} \\
\end{aligned}
\end{aligned}
\]
Similar formulations can be derived regarding parameters $p_{ij}^D$ and $p_{jk}^S$. Therefore, the ELBO is explicit regarding each variational parameter and the gradient of the ELBO $J(q)$ depending on the variational parameters $\mb a_{ik}$, $\mb b_{jk}$, $r_{ijk}$, $p_{ij}^D$ and $p_{jk}^S$ respectively can be derived to find the coordinate of the stationary point (corresponding to a local optimum). In our factor model all full conditionals are tractable (c.f. \cref{supp:sec:meth:conditional} in Supp. Mat.). In practice, thanks to the formulation in the exponential family, the optimum value for each variational parameter corresponds to the expectation regarding $q$ of the corresponding parameter of the full conditional distribution \cite[see][]{hoffman2013}. Thus the coordinates of the ELBO's gradient optimal point are explicit. We mentioned that distributions with a mass at 0 (zero-inflated Poisson or sparse Gamma) lie in the exponential family \citep{eggers2015} and the general formulation from \cite{hoffman2013} remains valid. Detailed formulations of update rules regarding all variational parameters are given in Supp. Mat. (\cref{supp:sec:meth:var_update}).

\subsection{Variational EM algorithm}\label{sec:meth:prior_update}

We use the variational-EM algorithm \citep{beal2003} to jointly approximate the posterior distributions and to estimate the hyper-parameters $\mbg\Omega = (\mbg\alpha, \mbg\beta, \mbg\pi^S, \mbg\pi^D)$. In this framework, the variational inference is used within a variational E-step, in which the standard expectation of the joint likelihood regarding the posterior $\EE[p(\mb X, \mb U, \mb V',\mb S, \mb D\,;\,\mbg\Omega)|\mb X]$ is approximated by $\EE_q[p(\mb X, \mb U, \mb V', \mb S, \mb D\,;\,\mbg\Omega)]$. Then the variational M-step consists in maximizing $\EE_q[p(\mb X, \mb U, \mb V', \mb S, \mb D\,;\,\mbg\Omega)]$ w.r.t. the hyper-parameters $\mbg\Omega$. In the variational-EM algorithm, we have explicit formulations of the stationary points regarding variational parameters (E-step) and prior hyper-parameters (M-step) in the model, thus we use a coordinate descent iterative algorithm \citep[see][for a review]{wright2015} to infer the variational distribution. Detailed formulations of update rules regarding all prior hyper-parameters are given in Supp. Mat. (\cref{supp:sec:meth:prior_update}).

\subsection{Initialization of the algorithm}\label{sec:meth:init}

\revise{}{To initialize variational and hyper-parameters of the model, we sample $\mb U$ and $\mb V$  from Gamma distributions such that  $X_{ij}\simeq \sum_k U_{ik}V_{jk}$ on average. The Gamma variational and hyper parameters are initialized from these values following the update rules detailed in \supp \cref{supp:sec:meth:var_update}. Dropout probabilities $p_{ij}^\text{D}$ and $\pi_{j}^\text{D}$ are initialized by $1/n \sum_i \II_{\{X_{ij}>0\}} $, i.e. the proportion of non-zero in the data for the corresponding gene. To initialize the sparsity probabilities $p_{jk}^\text{S}$ and $\pi_{j}^\text{S}$, we use a heuristic based on the variance of the corresponding gene $j$. Assuming that genes with low variability will have less impact on the structure embedded in the data, we propose a starting value such that 
\begin{equation}\label{eq:meth:init:sparse}
P_{j}^{(0)} = 1-\exp \left(- \widehat{s}_j / \widehat{m}_j \right)\, ,
\end{equation}
where $\widehat{m}_j$ is the mean of the non-null observations for gene $j$ and $ \widehat{s}_j$ its standard deviation (including null values). The scaling is better when removing the null values to compute the mean. This quantity adapts to the empirical variance of the observations, and will be close to 0 for genes with low variance, and close to 1 for genes with high variability.}

\section{Empirical study of pCMF}\label{sec:results}

\revise{}{All codes are available on a public repository for reproducibility\footnote{\url{https://github.com/gdurif/pCMF_experiments}}. We compare our method with standard approaches for unsupervised dimension reduction: the Poisson-NMF \citep{lee1999}, applied to raw counts (model-based matrix factorization approach based on the Poisson distribution); the PCA \citep{pearson1901} and  the sparse PCA \citep{witten2009}, based on an $\ell_1$ penalty in the optimization problem defining the PCA to induce sparsity in the loadings $\mb V$, both applied to log counts. We use sparse methods (sparse PCA, sparse pCMF) with a re-estimation step on the selected genes. We will refer to them as (s)PCA and (s)pCMF in the results respectively, to differentiate them from sparse PCA and sparse pCMF (without re-estimation), PCA and pCMF (without the sparsity layer).} In addition, we use the Zero-Inflated Factor Analysis (ZIFA) by \cite{pierson2015}, a dimension reduction approach that is specifically designed to handle dropout events in single-cell expression data (based on a zero-inflated Gaussian factor model applied to log-transformed counts). We present quantitative clustering results and qualitative visualization results on simulated and experimental scRNA-seq data. We also compare our method with t-SNE that is commonly used for data visualization \citep{vandermaaten2008}. It requires to choose a ``perplexity'' hyper-parameter that cannot be automatically calibrated, thus being less appropriate for a quantitative analysis. \revise{}{In the following, we always choose the perplexity values that gives the better clustering results.}

\subsection{Simulated data analysis}
To generate synthetic data we follow the hierarchical Gamma-Poisson framework as adopted by others \citep{zappia2017}. Details are provided in the Supp. Mat. (\cref{supp:sec:data_gen}). We generate synthetic multivariate over-dispersed counts, with $n=100$ individuals and \revise{$m=1000$}{$m=800$} recorded variables. We artificially create clusters of individuals \revise{}{(with different level of dispersion)} and groups of dependent variables. We we set different levels of zero-inflation in the data (i.e. low or high probabilities of dropout events, corresponding to random null values in the data), and some part of the $m$ variables are generated as random noise that do not induce any latent structure. Thus, we can test the performance of our method in different realistic data configurations, the range of our simulation parameters being comparable to other published simulated data \citep{zappia2017}.

\revise{}{We train the different methods with $K=2$ to visualize the reconstructed matrices $\hat{\mb U}$ and $\hat{\mb V}$ (c.f. \cref{sec:method}). To assess the ability of each method to retrieve the structure of cells or genes, we run a $\kappa$-mean clustering algorithm on $\hat{\mb U}$ and $\hat{\mb V}$ respectively (with $\kappa=2$) and we measure the adjusted Rand Index \citep{rand1971} quantifying the accordance between the predicted clusters and original groups of cells or genes. Regarding our approach pCMF, we use $\log \hat{\mb U}$ and $\log \hat{\mb V}$ for data visualization and clustering because the $\log$ is the canonical link function for Gamma models. In addition, we also compute the percentage of explained deviance associated to the model to assess the quality of the reconstruction. Regarding the PCA (sparse or not) and ZIFA, we use the standard explained variance criterion (c.f. \cref{sec:method:expdev}).}

\subsubsection{Clustering in the observation space}\label{sec:simu:cluster}

\paragraph{Effect of zero-inflation and cell representation.} 

\begin{figure}[!tpb]
\centering
\begin{subfigure}{.4\linewidth}
\includegraphics[width=.95\linewidth]{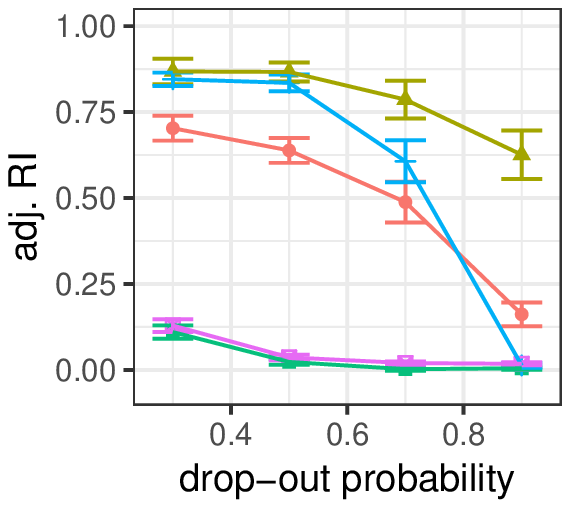}
\caption{}
\label{fig:simu:quant:3groups:adjri:U}
\end{subfigure}
\begin{subfigure}{0.58\linewidth}
\includegraphics[width=.94\linewidth]{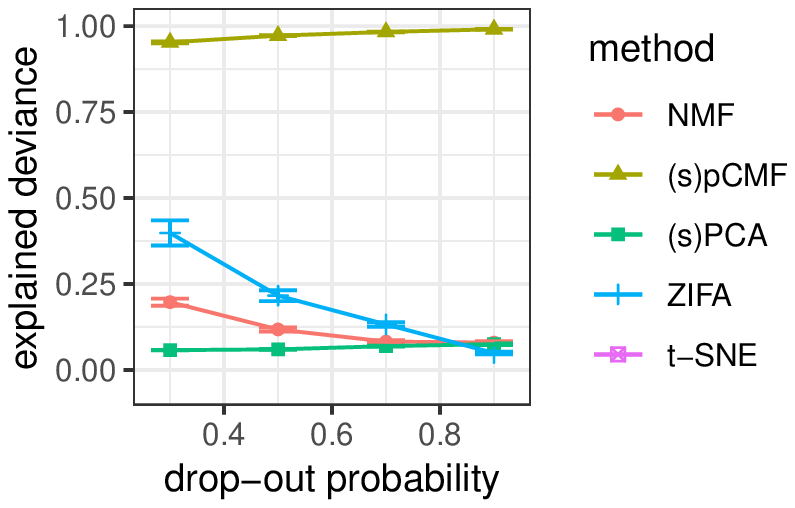}
\caption{}
\label{fig:simu:quant:3groups:expdev:U}
\end{subfigure}
\caption{Adjusted Rand Index (\ref{fig:simu:quant:3groups:adjri:U}) for the clustering on $\hat{\mb U}$ versus the true groups of cells; and explained deviance (\ref{fig:simu:quant:3groups:expdev:U}) depending on the probability used to generate dropout events. Average values and deviation are estimated across 50 repetitions.}
\label{fig:simu:quant:3groups:U}
\end{figure}

\revise{}{
We first assess the robustness of the different methods to the level of dropout or zero-inflation (ZI) in the data. We generate data with 3 groups of observations (c.f. Supp. Mat. \cref{supp:sec:data_gen}) with a wide range of dropout probabilities. \cref{fig:simu:quant:3groups:adjri:U} shows that (s)pCMF adapts to the level of dropout in the data and recovers the original clusters (high adjusted Rand Index) even with dropouts. Despite comparable performance with low dropout, Poisson-NMF and ZIFA are very sensitive to the addition of zeros. In addition, methods based on transformed counts like (s)PCA and t-SNE perform poorly, as they do not account for the specificity of the data (discrete, over-dispersed, \citep{ohara2010}).}

\begin{figure}[!tpb]
\centering
\begin{subfigure}{.4\linewidth}
\includegraphics[width=.95\linewidth]{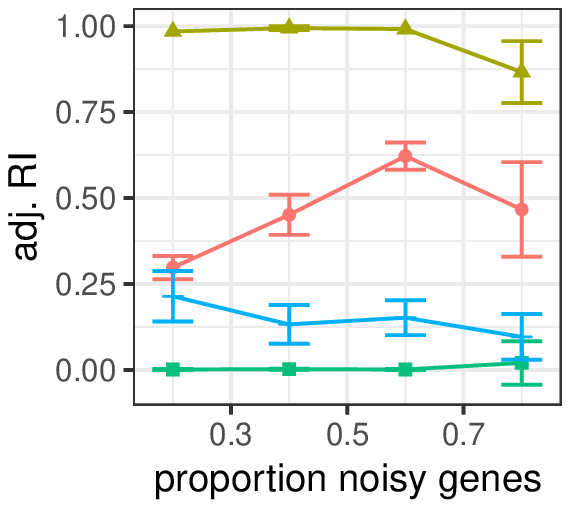}
\caption{}
\label{fig:simu:quant:3groups:adjri:V}
\end{subfigure}
\begin{subfigure}{0.58\linewidth}
\includegraphics[width=.94\linewidth]{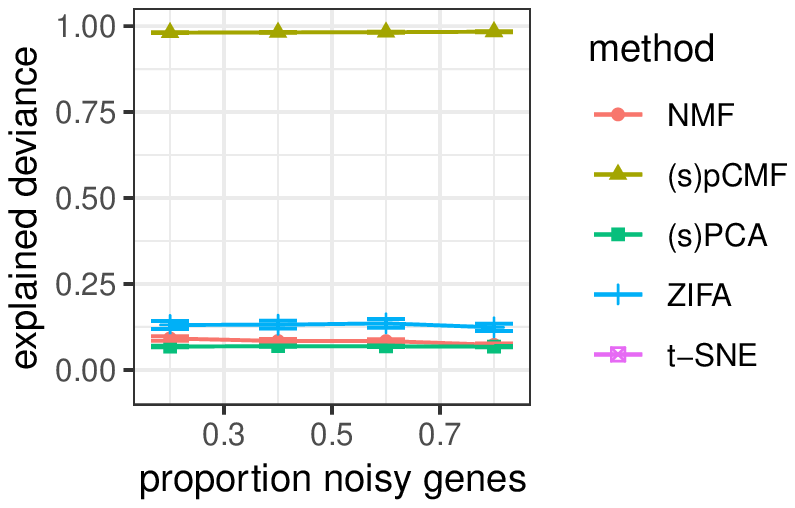}
\caption{}
\label{fig:simu:quant:3groups:expdev:V}
\end{subfigure}
\caption{Adjusted Rand Index (\ref{fig:simu:quant:3groups:adjri:V}) for the clustering on $\hat{\mb V}$ versus the true groups of genes; and explained deviance (\ref{fig:simu:quant:3groups:expdev:V}) depending on the proportion of noisy genes. Average values and deviation are estimated across 50 repetitions.}
\label{fig:simu:quant:3groups:V}
\end{figure}

\paragraph{Effect of noisy genes and gene representation.} \revise{}{To quantify the impact of noisy genes on the retrieval of the clusters, we consider data generated with different proportions of noisy genes (genes that do not induce any structure in the data). We generate data with three groups of genes: two groups inducing some latent structure and one group of noisy genes (c.f. Supp. Mat. \cref{supp:sec:data_gen}). \cref{fig:simu:quant:3groups:V} shows that (s)pCMF identifies correctly the clusters of genes, including the set of noisy genes, contrary to other approaches. This point shows that our approach correctly identifies the genes that support the lower-dimensional representation.}

\revise{}{In addition to the clustering results, we compared the selection accuracy of the only two methods (sPCA, spCMF) that perform variable selection (\supp \cref{supp:fig:simu:quant:3groups:accsel}). A selected gene is a gene that contributes to any latent dimension (any $V_{jk} \neq 0)$.  Sparse pCMF performs better than sparse PCA for various latent dimension $K$ even for high levels of noisy genes. Sparse PCA shows better selection accuracy when the proportion of noisy genes is low. This point suggests that sparse pCMF would be less sensitive to gene pre-filtering when analysing scRNA-seq data, which corresponds to a removal of noisy genes and is generally crucial \citep{soneson2018}.}

\revise{}{Details about data generation and additional data configuration regarding \cref{fig:simu:quant:3groups:U,fig:simu:quant:3groups:V} can be found in \supp (\cref{supp:sec:simu:results}, especially \cref{supp:fig:simu:quant:3groups:U,supp:fig:simu:quant:3groups:V}).}

\subsubsection{Data visualization}\label{sec:simu:visu}

\revise{}{
Data visualization is central in single-cell transcriptomics for the representation of high dimensional data in a lower dimensional space, in order to identify groups of cells, or to illustrate the cells diversity \citep[e.g.][]{llorens-bobadilla2015, segerstolpe2016}. In the matrix factorization framework, we represent observation (cell) coordinates and variable (gene) contributions: resp. $(\hat{u}_{i1},\hat{u}_{i2})_{i=1,\hdots,n}$ and $(\hat{v}_{i1},\hat{v}_{i2})_{i=1,\hdots,n}$ (or their $\log$ transform for pCMF) when the dimension is $K=2$ (see \cref{sec:method}).}

\begin{figure}[!tpb]
\centering
\includegraphics[width=.95\linewidth]{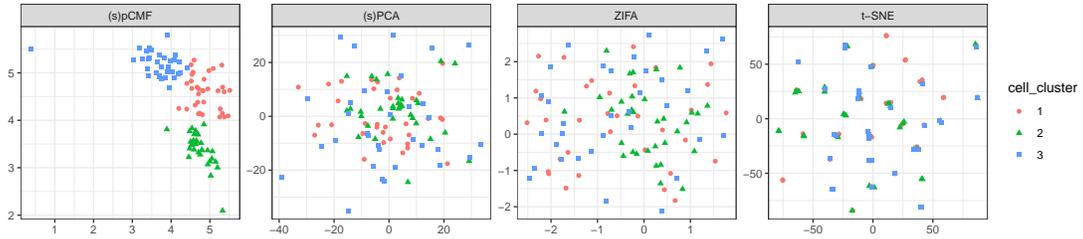}
\caption{Representation of cells in a subspace of dimension $K=2$. Here we have $60\%$ of noisy variables, and dropout probabilities around $0.9$.}
\label{fig:simu:visu:cells:3groups}
\end{figure}

\revise{}{We consider the same simulated data as previously ($n=100$, $m=800$, with three groups of cells, two groups of relevant genes and the set of noisy genes). Our visual results are consistent with the previous clustering results both regarding cell and gene visualization. In this challenging context (high zero-inflation and numerous noisy variables), by using pCMF, we are able to graphically identify the groups of individuals (cells) in the simulated zero-inflated count data (c.f. \cref{fig:simu:visu:cells:3groups}). On the contrary, the 2-D visualization is not successful with  PCA, ZIFA, Poisson-NMF and t-SNE, illustrating the interest of our data-specific approach. This point supports our claim that using data-specific model improves the quality of the reconstruction in the latent space.}

\begin{figure}[!tpb]
\centering
\includegraphics[width=.95\linewidth]{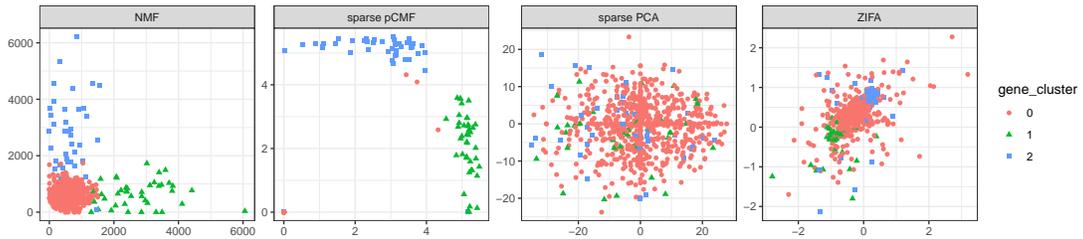}
\caption{Representation of genes in a subspace of dimension $K=2$ Here we have $80\%$ of noisy variables, and dropout probabilities around $0.7$. The label 0 corresponds to noisy genes.}
\label{fig:simu:visu:genes:3groups}
\end{figure}

\revise{}{In addition, linear projection methods (NMF, PCA, pCMF, ZIFA) can be used to visualize the contribution of genes to the principal axes (c.f. \cref{fig:simu:visu:genes:3groups}). Thanks to sparsity constraints,  the contribution of noisy genes are mostly set to 0 for sparse pCMF. Surprisingly, this selection is not efficient in the case of sparse PCA, indicating a lack of calibration of the sparsity constraint when data are counts. In comparison, Poisson NMF and ZIFA (not sparse) do not identity the cluster of noisy genes as clearly as  sparse pCMF.}

\revise{}{To quantify the model quality of the different methods on simulated data, we used the deviance associated to each method (c.f. \cref{sec:method:expdev}). \cref{fig:simu:quant:3groups:expdev:U,fig:simu:quant:3groups:expdev:V}  shows that the dimension reduction proposed by pCMF has excellent fit to the data regardless the level of dropout or the proportion of noisy genes, as compared with other methods. Additional comparisons of computational time show that PCA is fastest method (but with low performance), whereas (sparse) pCMF is faster than ZIFA and sparse PCA, with increased performance (cf Supp. Mat. \cref{supp:sec:simu:results}).} 

\subsection{Analysis of single-cell data}\label{sec:data}

\revise{}{We now illustrate the performance of pCMF on different recent and large single-cell RNA-seq datasets that are publicly availbale: the \cite{baron2016} dataset, the goldstandard and silverstandard datasets used in \cite{freytag2018} (we used the silverstandard dataset 5 which was the largest). We also consider an older and smaller dataset from \citep{llorens-bobadilla2015} which is interesting because it discribes a continuum of activation in Neural Stem Cells (NSC). All datasets are available here\footnote{\url{https://github.com/gdurif/pCMF_experiments}} with the codes. More details about their origins are given in \supp \cref{supp:sec:data}. We consider large datasets with $\sim 1000$ or $\sim 10000$ cells to test the ability of our approach to face the expected increase of data volume in the next couple of years.}

\begin{figure}[h]
\centering
\includegraphics[width=.99\linewidth]{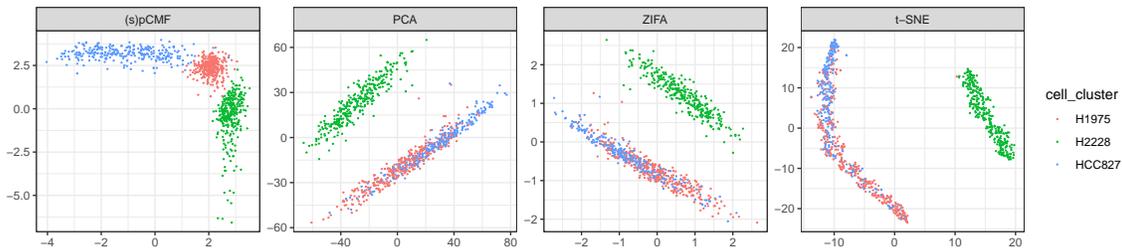}
\caption{Analysis of the goldstandard scRNA-seq data from \cite{freytag2018}, $925$ cells, $8580$ genes. Visualization of the cells in a latent space of dimension 2.}
\label{fig:data:visu:gold:cells}
\end{figure}

\revise{}{We present some quantitative results about clustering and data reconstruction (c.f. \cref{table:data:criteria}) and the corresponding qualitative results about cell visualization (c.f. \cref{fig:data:visu:gold:cells,supp:fig:data:visu:baron:cells,supp:fig:data:visu:silver:cells,supp:fig:data:visu:llorens:cells} in Supp. Mat.) and gene visualization  (c.f. \cref{supp:fig:data:visu:baron:genes,supp:fig:data:visu:gold:genes,supp:fig:data:visu:silver:genes,supp:fig:data:visu:llorens:genes} in Supp. Mat.). For each dataset \cite[except the one from][where we used their pre-filtering]{llorens-bobadilla2015}, we use the same pipeline, we filter out genes expressed in less than $5\%$ of the cells. In a second step, we remove genes for which the variance heuristic defined in \cref{eq:meth:init:sparse} is low. In practice we removed genes for which $P_j^{(0)} \le 0.2$. Our idea was to remove uninformative genes, since pre-filtering is crucial \citep{soneson2018} in such data, but also to reduce the number of genes to reduce the computation cost, in particular for ZIFA. Then we compare (s)pCMF, PCA, ZIFA and t-SNE. We discarded (s)PCA because the sparse PCA is computationally expensive (c.f. \supp \cref{supp:sec:simu:time}) due to the required cross-validation.}

\revise{}{A general empirical property is that clustering accuracy dicreases for all methods when the number of groups of cells increases. However, our approach (s)pCMF produces a better (or as good) view of the cells regarding clustering purpose in every examples, since the adjusted Rand Index is higher (c.f. \cref{table:data:criteria}). We observe the same trend regarding the quality of the reconstruction since the explained deviance is generally also higher for (s)pCMF. Data visualization is not always clear (c.f. \cref{supp:fig:data:visu:baron:cells,supp:fig:data:visu:silver:cells}), especially when the number of groups is large as in \cite{baron2016} or \cite{freytag2018} silverstandard, however it is possible to clearly identify large clusters of cells in the (e.g. beta cells in \cite{baron2016} or CD14+ Monocyte in \cite{freytag2018} silverstandard) with our method (and some of the others). On the goldstandard dataset from \cite{freytag2018}, the difference regarding cells representation between the different approaches is more visual (c.f. \cref{fig:data:visu:gold:cells}), where our approach is the only one that is able to distinctly identify the three populations of cells. On the \cite{llorens-bobadilla2015} dataset, our approach clearly highlights this continuum of activation presented in their paper, which can also be seen with ZIFA, but is not as much clear with PCA and t-SNE.}

\revise{}{Regarding gene visualization (c.f. \cref{supp:fig:data:visu:baron:genes,supp:fig:data:visu:gold:genes,supp:fig:data:visu:silver:genes,supp:fig:data:visu:llorens:genes} in Supp. Mat.), we compare the representation of sparse pCMF to PCA, ZIFA (and not t-SNE since it does not jointly learn $\mb U$ and $\mb V$). The interest of sparsity for gene selection is to highlight more precisely the genes that contribute to the latent representation. For each dataset, it is possible to detect which genes are important for each latent dimension: some are null on both (e.g. uninformative genes), some contribute to a single dimension, some contribute to both. This pattern is not as clear with methods that do not implement any sparsity layers.}

\revise{}{These different points show the interest of our approach to analyze recent single-cell RNA-seq datasets, even large ones. Empirical properties studied on simulations are confirmed on experimental data: providing a dimension reduction method adapted to single cell data, where the sparsity constraints is powerful to represent complex single cell data. In addition, our heuristic to assume gene importance based on their variance appears to be efficient $(i)$ to perform a rough pre-filtering to reduce the dimension and $(ii)$ to discriminate between noisy genes and informative ones directly in the sparse pCMF algorithm. In addition, it appears that our method is fast compared to ZIFA for instance, since it takes less than two minutes on the different examples to run (s)pCMF (sparse pCMF + re-estimation), on a 16-core machine, whereas ZIFA can take between 5 and 25 minutes on the same architecture.}

\FloatBarrier
\afterpage{
\clearpage
\begin{landscape}
\begin{table}
	\begin{tabular}{@{}lccccclcccccc@{}}
		& nb & nb genes & \multirow{2}{*}{prop. 0} & nb & & & \multirow{2}{*}{(s)pCMF} & \multirow{2}{*}{PCA} & \multirow{2}{*}{ZIFA} & \multirow{2}{*}{t-SNE} \\
		& cells & (before pre-filter.) & & group & & & & & & \\
		\hline
		\multirow{2}{*}{\cite{baron2016}} & \multirow{2}{*}{1886} & 6080 & \multirow{2}{*}{$80.9\%$} & \multirow{2}{*}{13} & & adj. RI & $\mb{21.2\%}$ & $14.3\%$ & $15.4\%$ & $14.2\%$ \\
		& & (14878) & & & & \%dev & $\mb{73.2\%}$ & $41.6\%$ & $53.5\%$ &  / \\
		\hline
		\cite{freytag2018} & \multirow{2}{*}{925} & 8580 & \multirow{2}{*}{$39.5\%$} & \multirow{2}{*}{3} & & adj. RI & $\mb{81.3\%}$ & $60.1\%$ & $56.8\%$ & $60.5\%$ \\
		goldstandard & & (58302) & & & & \%dev & $55.7\%$ & $\mb{65.6\%}$ & $48.6\%$ & / \\
		\hline
		\cite{freytag2018} & \multirow{2}{*}{8352} & 4547 & \multirow{2}{*}{$86.3\%$} & \multirow{2}{*}{11} & & adj. RI & $24.2\%$ & $16.2\%$ & $19.8\%$ & $\mb{24.8\%}$ \\
		silverstandard 5 & & (33694) & & & & \%dev & $\mb{70.0\%}$ & $55.1\%$ & / & / \\
		\hline
		\multirow{2}{*}{\cite{llorens-bobadilla2015}} & \multirow{2}{*}{141} & 13826 & \multirow{2}{*}{$64.8\%$} & \multirow{2}{*}{6} & & adj. RI & $\mb{40.1\%}$ & $25.3\%$ & $38.3\%$ & $29.8\%$ \\
		& & (43309) & & & & \%dev & $\mb{64.4\%}$ & $34.8\%$ & $42.6\%$ & / \\
	\end{tabular}
\caption{Performance of the different methods regarding quality of reconstruction (percentage of explained deviance) and clustering (adjusted Rand Index). Each scRNA-seq dataset is characterized by the number of cells, the number of genes used in the analysis (we specify the original number before the pre-filtering step) and by the number of original groups. The adjusted Rand Index compares clusters found by a $\kappa$-means algorithm (applied to $\hat{\mb U}$ with $\kappa=$ nb group) and the original groups of cells.\label{table:data:criteria}}
\end{table}
\end{landscape}
\clearpage
}

\FloatBarrier

\newpage

\section{Conclusion}

In this work, we provide a new framework for dimension reduction in unsupervised context. In particular, we introduce a model-based matrix factorization method specifically designed to analyze single-cell RNA-seq data. Matrix factorization allows to jointly construct a lower dimensional representation of cells and genes. Our probabilistic Count Matrix Factorization (pCMF) approach accounts for the specificity of these data, being zero-inflated and over-dispersed counts. In other words, we propose a generalized PCA procedure that is suitable for data visualization and clustering. The interest of our zero-inflated sparse Gamma-Poisson factor model is to replace the variance-based formulation of PCA, associated to the Euclidean geometry and the Gaussian distribution, with a metric (based on Bregman divergence) that is adapted to scRNA-seq data characteristics.

Analyzing single-cell expression profiles is a huge challenge to understand the cell diversity in a tissue/an organism and more precisely for characterizing the associated gene activity. We show on simulations and experimental data that our pCMF approach is able to catch the underlying structure in zero-inflated over-dispersed count data. In particular, we show that our method can be used for data visualization in a lower dimensional space or for preliminary dimension reduction before a clustering step. In both cases, pCMF performs as well or out-performs state-of-the-art approaches, especially the PCA (being the gold standard) or more specific methods such as the NMF (count based) or ZIFA (zero-inflation specific). In particular, pCMF data representation is less sensitive to the choice of the latent dimension $K$ regarding clustering results, which supports the interest of our approach for data exploration. It appears (through the explained deviance criterion that we introduced) that the reconstruction learned by pCMF better represents the variability in the data (compared to PCA or ZIFA). In addition, \revise{}{pCMF can select genes that explain the latent structure in the data, thanks to a sparsity layer which does not require any parameter tuning.}

\revise{}{An interesting direction to improve pCMF would be to integrate covariables or confounding factors in the Gamma-Poisson model, for instance to account for technical effect in the data or for data normalization. A similar framework based on zero-inflated Negative Binomial distribution was proposed by \cite{risso2017}, and it could be extended to our framework of matrix factorization}


\section*{Funding}

This work was supported by the french National Resarch Agency (ANR) as part of the ``ABS4NGS'' project [ANR-11-BINF-0001-06] and as part of the ``MACARON'' project [ANR-14-CE23-0003], and by the European Research Council as part of the ERC grant Solaris. It was performed using the computing facilities of the computing center LBBE/PRABI.

\bibliographystyle{natbib}
\bibliography{article_pCMF_2019}

\begin{thebibliography}{}

\bibitem[Amir {\em et~al.}(2013)Amir, Davis, Tadmor, Simonds, Levine, Bendall,
  Shenfeld, Krishnaswamy, Nolan, and Pe'er]{ADT13}
Amir, e.-A., Davis, K.~L., Tadmor, M.~D., Simonds, E.~F., Levine, J.~H.,
  Bendall, S.~C., Shenfeld, D.~K., Krishnaswamy, S., Nolan, G.~P., and Pe'er,
  D. (2013).
\newblock {vi{S}{N}{E} enables visualization of high dimensional single-cell
  data and reveals phenotypic heterogeneity of leukemia}.
\newblock {\em Nat. Biotechnol.}, {\bf 31}(6), 545--552.

\bibitem[Anders and Huber(2010)Anders and Huber]{anders2010}
Anders, S. and Huber, W. (2010).
\newblock Differential expression analysis for sequence count data.
\newblock {\em Genome Biology\/}, {\bf 11}(10), R106.

\bibitem[Banerjee {\em et~al.}(2005)Banerjee, Merugu, Dhillon, and
  Ghosh]{banerjee2005}
Banerjee, A., Merugu, S., Dhillon, I.~S., and Ghosh, J. (2005).
\newblock Clustering with {{Bregman Divergences}}.
\newblock {\em Journal of Machine Learning Research\/}, {\bf 6}(Oct),
  1705--1749.

\bibitem[Baron {\em et~al.}(2016)Baron, Veres, Wolock, Faust, Gaujoux, Vetere,
  Ryu, Wagner, {Shen-Orr}, Klein, Melton, and Yanai]{baron2016}
Baron, M., Veres, A., Wolock, S.~L., Faust, A.~L., Gaujoux, R., Vetere, A.,
  Ryu, J.~H., Wagner, B.~K., {Shen-Orr}, S.~S., Klein, A.~M., Melton, D.~A.,
  and Yanai, I. (2016).
\newblock A {{Single}}-{{Cell Transcriptomic Map}} of the {{Human}} and {{Mouse
  Pancreas Reveals Inter}}- and {{Intra}}-cell {{Population Structure}}.
\newblock {\em Cell systems\/}, {\bf 3}(4), 346--360.e4.

\bibitem[Beal and Ghahramani(2003)Beal and Ghahramani]{beal2003}
Beal, M.~J. and Ghahramani, Z. (2003).
\newblock The variational {{Bayesian EM}} algorithm for incomplete data: With
  application to scoring graphical model structures.
\newblock {\em Bayesian statistics\/}, {\bf 7}, 453--464.

\bibitem[Blei {\em et~al.}(2017)Blei, Kucukelbir, and McAuliffe]{blei2017}
Blei, D.~M., Kucukelbir, A., and McAuliffe, J.~D. (2017).
\newblock Variational inference: {{A}} review for statisticians.
\newblock {\em Journal of the American Statistical Association\/},
  (just-accepted).

\bibitem[Buenrostro {\em et~al.}(2015)Buenrostro, Wu, Litzenburger, Ruff,
  Gonzales, Snyder, Chang, and Greenleaf]{BWL15}
Buenrostro, J.~D., Wu, B., Litzenburger, U.~M., Ruff, D., Gonzales, M.~L.,
  Snyder, M.~P., Chang, H.~Y., and Greenleaf, W.~J. (2015).
\newblock {{S}ingle-cell chromatin accessibility reveals principles of
  regulatory variation}.
\newblock {\em Nature\/}, {\bf 523}(7561), 486--490.

\bibitem[Buettner {\em et~al.}(2015)Buettner, Natarajan, Casale, Proserpio,
  Scialdone, Theis, Teichmann, Marioni, and Stegle]{BNC15}
Buettner, F., Natarajan, K.~N., Casale, F.~P., Proserpio, V., Scialdone, A.,
  Theis, F.~J., Teichmann, S.~A., Marioni, J.~C., and Stegle, O. (2015).
\newblock {{C}omputational analysis of cell-to-cell heterogeneity in
  single-cell {R}{N}{A}-sequencing data reveals hidden subpopulations of
  cells}.
\newblock {\em Nat. Biotechnol.}, {\bf 33}(2), 155--160.

\bibitem[Cemgil(2009)Cemgil]{cemgil2009}
Cemgil, A.~T. (2009).
\newblock Bayesian inference for nonnegative matrix factorisation models.
\newblock {\em Computational Intelligence and Neuroscience\/}, {\bf 2009}.

\bibitem[Chen {\em et~al.}(2016)Chen, Jin, Huang, and Chen]{chen2016}
Chen, H.-I.~H., Jin, Y., Huang, Y., and Chen, Y. (2016).
\newblock Detection of high variability in gene expression from single-cell
  {{RNA}}-seq profiling.
\newblock {\em BMC Genomics\/}, {\bf 17}(Suppl 7).

\bibitem[Chen {\em et~al.}(2008)Chen, Chen, and Rao]{chen2008}
Chen, P., Chen, Y., and Rao, M. (2008).
\newblock Metrics defined by {{Bregman Divergences}}.
\newblock {\em Communications in Mathematical Sciences\/}, {\bf 6}(4),
  915--926.

\bibitem[Collins {\em et~al.}(2001)Collins, Dasgupta, and
  Schapire]{collins2001}
Collins, M., Dasgupta, S., and Schapire, R.~E. (2001).
\newblock A generalization of principal components analysis to the exponential
  family.
\newblock In {\em Advances in Neural Information Processing Systems\/}, pages
  617--624.

\bibitem[Deng {\em et~al.}(2014)Deng, Ramsk{\"o}ld, Reinius, and
  Sandberg]{deng2014}
Deng, Q., Ramsk{\"o}ld, D., Reinius, B., and Sandberg, R. (2014).
\newblock Single-cell {{RNA}}-seq reveals dynamic, random monoallelic gene
  expression in mammalian cells.
\newblock {\em Science\/}, {\bf 343}(6167), 193--196.

\bibitem[Dikmen and F{\'e}votte(2012)Dikmen and F{\'e}votte]{dikmen2012}
Dikmen, O. and F{\'e}votte, C. (2012).
\newblock Maximum marginal likelihood estimation for nonnegative dictionary
  learning in the {{Gamma}}-{{Poisson}} model.
\newblock {\em Signal Processing, IEEE Transactions on\/}, {\bf 60}(10),
  5163--5175.

\bibitem[Eckart and Young(1936)Eckart and Young]{eckart1936}
Eckart, C. and Young, G. (1936).
\newblock The approximation of one matrix by another of lower rank.
\newblock {\em Psychometrika\/}, {\bf 1}(3), 211--218.

\bibitem[Eggers(2015)Eggers]{eggers2015}
Eggers, J. (2015).
\newblock On {{Statistical Methods}} for {{Zero}}-{{Inflated Models}}.
\newblock Technical Report U.U.D.M. Project Report 2015:9, Uppsala Universitet.

\bibitem[Engelhardt and Adams(2014)Engelhardt and Adams]{engelhardt2014}
Engelhardt, B.~E. and Adams, R.~P. (2014).
\newblock Bayesian {{Structured Sparsity}} from {{Gaussian Fields}}.
\newblock {\em arXiv:1407.2235 [q-bio, stat]\/}.

\bibitem[F{\'e}votte and Cemgil(2009)F{\'e}votte and Cemgil]{fevotte2009}
F{\'e}votte, C. and Cemgil, A.~T. (2009).
\newblock Nonnegative matrix factorizations as probabilistic inference in
  composite models.
\newblock In {\em Signal {{Processing Conference}}, 2009 17th {{European}}\/},
  pages 1913--1917. {IEEE}.

\bibitem[Fraley and Raftery(2002)Fraley and Raftery]{fraley2002}
Fraley, C. and Raftery, A.~E. (2002).
\newblock Model-based clustering, discriminant analysis, and density
  estimation.
\newblock {\em Journal of the American statistical Association\/}, {\bf
  97}(458), 611--631.

\bibitem[Freytag {\em et~al.}(2018)Freytag, Tian, L\"onnstedt, Ng, and
  Bahlo]{freytag2018}
Freytag, S., Tian, L., L\"onnstedt, I., Ng, M., and Bahlo, M. (2018).
\newblock Comparison of clustering tools in {{R}} for medium-sized 10x
  {{Genomics}} single-cell {{RNA}}-sequencing data.
\newblock {\em F1000Research\/}, {\bf 7}.

\bibitem[Friguet(2010)Friguet]{friguet2010}
Friguet, C. (2010).
\newblock {\em Impact de La D{\'e}pendance Dans Les Proc{\'e}dures de Tests
  Multiples En Grande Dimension\/}.
\newblock Ph.D. thesis, Rennes, AGROCAMPUS-OUEST.

\bibitem[Gaujoux and Seoighe(2010)Gaujoux and Seoighe]{gaujoux2010}
Gaujoux, R. and Seoighe, C. (2010).
\newblock A flexible {{R}} package for nonnegative matrix factorization.
\newblock {\em BMC Bioinformatics\/}, {\bf 11}, 367.

\bibitem[Golub {\em et~al.}(1999)Golub, Slonim, Tamayo, Huard, Gaasenbeek,
  Mesirov, Coller, Loh, Downing, Caligiuri, Bloomfield, and Lander]{GST99}
Golub, T.~R., Slonim, D.~K., Tamayo, P., Huard, C., Gaasenbeek, M., Mesirov,
  J.~P., Coller, H., Loh, M.~L., Downing, J.~R., Caligiuri, M.~A., Bloomfield,
  C.~D., and Lander, E.~S. (1999).
\newblock {{M}olecular classification of cancer: class discovery and class
  prediction by gene expression monitoring}.
\newblock {\em Science\/}, {\bf 286}(5439), 531--537.

\bibitem[Gr{\"u}n {\em et~al.}(2015)Gr{\"u}n, Lyubimova, Kester, Wiebrands,
  Basak, Sasaki, Clevers, and {van Oudenaarden}]{grun2015}
Gr{\"u}n, D., Lyubimova, A., Kester, L., Wiebrands, K., Basak, O., Sasaki, N.,
  Clevers, H., and {van Oudenaarden}, A. (2015).
\newblock Single-cell messenger {{RNA}} sequencing reveals rare intestinal cell
  types.
\newblock {\em Nature\/}, {\bf 525}(7568), 251.

\bibitem[Hoffman {\em et~al.}(2013)Hoffman, Blei, Wang, and
  Paisley]{hoffman2013}
Hoffman, M.~D., Blei, D.~M., Wang, C., and Paisley, J. (2013).
\newblock Stochastic {{Variational Inference}}.
\newblock {\em J. Mach. Learn. Res.}, {\bf 14}(1), 1303--1347.

\bibitem[Kharchenko {\em et~al.}(2014)Kharchenko, Silberstein, and
  Scadden]{kharchenko2014}
Kharchenko, P.~V., Silberstein, L., and Scadden, D.~T. (2014).
\newblock Bayesian approach to single-cell differential expression analysis.
\newblock {\em Nature Methods\/}, {\bf 11}(7), 740.

\bibitem[Krijthe(2015)Krijthe]{krijthe2015}
Krijthe, J.~H. (2015).
\newblock {\em {Rtsne}: T-Distributed Stochastic Neighbor Embedding using
  Barnes-Hut Implementation\/}.
\newblock R package version 0.13.

\bibitem[Landgraf and Lee(2015)Landgraf and Lee]{landgraf2015}
Landgraf, A.~J. and Lee, Y. (2015).
\newblock Generalized principal component analysis: {{Projection}} of saturated
  model parameters.
\newblock {\em Technical Report 892, Department of Statistics, The Ohio State
  University\/}.

\bibitem[Lee and Seung(1999)Lee and Seung]{lee1999}
Lee, D.~D. and Seung, H.~S. (1999).
\newblock Learning the parts of objects by non-negative matrix factorization.
\newblock {\em Nature\/}, {\bf 401}(6755), 788--791.

\bibitem[Llorens-Bobadilla {\em et~al.}(2015)Llorens-Bobadilla, Zhao, Baser,
  Saiz-Castro, Zwadlo, and Martin-Villalba]{llorens-bobadilla2015}
Llorens-Bobadilla, E., Zhao, S., Baser, A., Saiz-Castro, G., Zwadlo, K., and
  Martin-Villalba, A. (2015).
\newblock Single-{{Cell Transcriptomics Reveals}} a {{Population}} of {{Dormant
  Neural Stem Cells}} that {{Become Activated}} upon {{Brain Injury}}.
\newblock {\em Cell Stem Cell\/}, {\bf 17}(3), 329--340.

\bibitem[Lun and Risso(2019)Lun and Risso]{lun2019}
Lun, A. and Risso, D. (2019).
\newblock {\em SingleCellExperiment: S4 Classes for Single Cell Data\/}.
\newblock R package version 1.4.1.

\bibitem[Malsiner-Walli and Wagner(2011)Malsiner-Walli and
  Wagner]{malsiner-walli2011}
Malsiner-Walli, G. and Wagner, H. (2011).
\newblock Comparing spike and slab priors for {{Bayesian}} variable selection.
\newblock {\em Austrian Journal of Statistics\/}, {\bf 40}(4), 241--264.

\bibitem[Minka(2000)Minka]{minka2000}
Minka, T. (2000).
\newblock Estimating a {{Dirichlet}} distribution.
\newblock Technical report, {MIT}.

\bibitem[Mitchell and Beauchamp(1988)Mitchell and Beauchamp]{mitchell1988}
Mitchell, T.~J. and Beauchamp, J.~J. (1988).
\newblock Bayesian variable selection in linear regression.
\newblock {\em Journal of the American Statistical Association\/}, {\bf
  83}(404), 1023--1032.

\bibitem[Nathoo {\em et~al.}(2013)Nathoo, Lesperance, Lawson, and
  Dean]{nathoo2013}
Nathoo, F.~S., Lesperance, M.~L., Lawson, A.~B., and Dean, C.~B. (2013).
\newblock Comparing variational {{Bayes}} with {{Markov}} chain {{Monte Carlo}}
  for {{Bayesian}} computation in neuroimaging.
\newblock {\em Statistical methods in medical research\/}, {\bf 22}(4),
  398--423.

\bibitem[O'Hara and Kotze(2010)O'Hara and Kotze]{ohara2010}
O'Hara, R.~B. and Kotze, D.~J. (2010).
\newblock Do not log-transform count data.
\newblock {\em Methods in Ecology and Evolution\/}, {\bf 1}(2), 118--122.

\bibitem[Pearson(1901)Pearson]{pearson1901}
Pearson, K. (1901).
\newblock {{LIII}}. {{On}} lines and planes of closest fit to systems of points
  in space.
\newblock {\em Philosophical Magazine Series 6\/}, {\bf 2}(11), 559--572.

\bibitem[Picelli {\em et~al.}(2013)Picelli, Bjorklund, Faridani, Sagasser,
  Winberg, and Sandberg]{PBF13}
Picelli, S., Bjorklund, A.~K., Faridani, O.~R., Sagasser, S., Winberg, G., and
  Sandberg, R. (2013).
\newblock {{S}mart-seq2 for sensitive full-length transcriptome profiling in
  single cells}.
\newblock {\em Nat. Methods\/}, {\bf 10}(11), 1096--1098.

\bibitem[Pierson and Yau(2015)Pierson and Yau]{pierson2015}
Pierson, E. and Yau, C. (2015).
\newblock {{ZIFA}}: {{Dimensionality}} reduction for zero-inflated single-cell
  gene expression analysis.
\newblock {\em Genome Biology\/}, {\bf 16}, 241.

\bibitem[Rand(1971)Rand]{rand1971}
Rand, W.~M. (1971).
\newblock Objective {{Criteria}} for the {{Evaluation}} of {{Clustering
  Methods}}.
\newblock {\em Journal of the American Statistical Association\/}, {\bf
  66}(336), 846--850.

\bibitem[Riggs and Lalonde(2017)Riggs and Lalonde]{riggs2017}
Riggs, J.~D. and Lalonde, T.~L. (2017).
\newblock {\em Handbook for {{Applied Modeling}}: {{Non}}-{{Gaussian}} and
  {{Correlated Data}}\/}.
\newblock {Cambridge University Press}.

\bibitem[Risso {\em et~al.}(2017)Risso, Perraudeau, Gribkova, Dudoit, and
  Vert]{risso2017}
Risso, D., Perraudeau, F., Gribkova, S., Dudoit, S., and Vert, J.-P. (2017).
\newblock {{ZINB}}-{{WaVE}}: {{A}} general and flexible method for signal
  extraction from single-cell {{RNA}}-seq data.
\newblock {\em bioRxiv\/}, page 125112.

\bibitem[Rotem {\em et~al.}(2015)Rotem, Ram, Shoresh, Sperling, Goren, Weitz,
  and Bernstein]{RRS15}
Rotem, A., Ram, O., Shoresh, N., Sperling, R.~A., Goren, A., Weitz, D.~A., and
  Bernstein, B.~E. (2015).
\newblock {{S}ingle-cell {C}h{I}{P}-seq reveals cell subpopulations defined by
  chromatin state}.
\newblock {\em Nat. Biotechnol.}, {\bf 33}(11), 1165--1172.

\bibitem[Saliba {\em et~al.}(2014)Saliba, Westermann, Gorski, and
  Vogel]{saliba2014}
Saliba, A.-E., Westermann, A.~J., Gorski, S.~A., and Vogel, J. (2014).
\newblock Single-cell {{RNA}}-seq: Advances and future challenges.
\newblock {\em Nucleic Acids Research\/}, {\bf 42}(14), 8845--8860.

\bibitem[Segerstolpe {\em et~al.}(2016)Segerstolpe, Palasantza, Eliasson,
  Andersson, Andr{\'e}asson, Sun, Picelli, Sabirsh, Clausen, Bjursell, Smith,
  Kasper, {\"A}mm{\"a}l{\"a}, and Sandberg]{segerstolpe2016}
Segerstolpe, {\AA}., Palasantza, A., Eliasson, P., Andersson, E.-M.,
  Andr{\'e}asson, A.-C., Sun, X., Picelli, S., Sabirsh, A., Clausen, M.,
  Bjursell, M.~K., Smith, D.~M., Kasper, M., {\"A}mm{\"a}l{\"a}, C., and
  Sandberg, R. (2016).
\newblock Single-{{Cell Transcriptome Profiling}} of {{Human Pancreatic
  Islets}} in {{Health}} and {{Type}} 2 {{Diabetes}}.
\newblock {\em Cell Metabolism\/}, {\bf 24}(4), 593--607.

\bibitem[Simchowitz(2013)Simchowitz]{simchowitz2013}
Simchowitz, M. (2013).
\newblock Zero-{{Inflated Poisson Factorization}} for {{Recommendation
  Systems}}.
\newblock {\em Junior Independent Work (advised by D. Blei), Princeton
  University, Department of Mathematics\/}.

\bibitem[Soneson and Robinson(2018)Soneson and Robinson]{soneson2018}
Soneson, C. and Robinson, M.~D. (2018).
\newblock Bias, robustness and scalability in single-cell differential
  expression analysis.
\newblock {\em Nature Methods\/}, {\bf 15}(4), 255--261.

\bibitem[Titsias and L{\'a}zaro-Gredilla(2011)Titsias and
  L{\'a}zaro-Gredilla]{titsias2011}
Titsias, M.~K. and L{\'a}zaro-Gredilla, M. (2011).
\newblock Spike and slab variational inference for multi-task and multiple
  kernel learning.
\newblock In {\em Advances in Neural Information Processing Systems\/}, pages
  2339--2347.

\bibitem[Trapnell {\em et~al.}(2014)Trapnell, Cacchiarelli, Grimsby, Pokharel,
  Li, Morse, Lennon, Livak, Mikkelsen, and Rinn]{TCG14}
Trapnell, C., Cacchiarelli, D., Grimsby, J., Pokharel, P., Li, S., Morse, M.,
  Lennon, N.~J., Livak, K.~J., Mikkelsen, T.~S., and Rinn, J.~L. (2014).
\newblock {{T}he dynamics and regulators of cell fate decisions are revealed by
  pseudotemporal ordering of single cells}.
\newblock {\em Nat. Biotechnol.}, {\bf 32}(4), 381--386.

\bibitem[{van der Maaten} and Hinton(2008){van der Maaten} and
  Hinton]{vandermaaten2008}
{van der Maaten}, L. and Hinton, G. (2008).
\newblock Visualizing {{Data}} using t-{{SNE}}.
\newblock {\em Journal of Machine Learning Research\/}, {\bf 9}(Nov),
  2579--2605.

\bibitem[Wagner {\em et~al.}(2016)Wagner, Regev, and Yosef]{WRY16}
Wagner, A., Regev, A., and Yosef, N. (2016).
\newblock {{R}evealing the vectors of cellular identity with single-cell
  genomics}.
\newblock {\em Nat. Biotechnol.}, {\bf 34}(11), 1145--1160.

\bibitem[Witten {\em et~al.}(2009)Witten, Tibshirani, and Hastie]{witten2009}
Witten, D.~M., Tibshirani, R., and Hastie, T. (2009).
\newblock A penalized matrix decomposition, with applications to sparse
  principal components and canonical correlation analysis.
\newblock {\em Biostatistics\/}, {\bf 10}(3), 515--534.

\bibitem[Wright(2015)Wright]{wright2015}
Wright, S.~J. (2015).
\newblock Coordinate descent algorithms.
\newblock {\em Mathematical Programming\/}, {\bf 151}(1), 3--34.

\bibitem[Zappia {\em et~al.}(2017)Zappia, Phipson, and Oshlack]{zappia2017}
Zappia, L., Phipson, B., and Oshlack, A. (2017).
\newblock Splatter: Simulation of single-cell {{RNA}} sequencing data.
\newblock {\em Genome Biology\/}, {\bf 18}, 174.

\bibitem[Zhou {\em et~al.}(2012)Zhou, Hannah, Dunson, and Carin]{zhou2012}
Zhou, M., Hannah, L.~A., Dunson, D.~B., and Carin, L. (2012).
\newblock Beta-negative binomial process and {{Poisson}} factor analysis.
\newblock In {\em In {{AISTATS}}\/}.

\bibitem[Zong {\em et~al.}(2012)Zong, Lu, Chapman, and Xie]{ZLC12}
Zong, C., Lu, S., Chapman, A.~R., and Xie, X.~S. (2012).
\newblock {{G}enome-wide detection of single-nucleotide and copy-number
  variations of a single human cell}.
\newblock {\em Science\/}, {\bf 338}(6114), 1622--1626.

\end{thebibliography}

\FloatBarrier
\newpage
\setcounter{section}{0}
\setcounter{figure}{0}
\setcounter{table}{0}
\setcounter{equation}{0}

\renewcommand{\thetable}{S.\arabic{table}}
\renewcommand{\thefigure}{S.\arabic{figure}}
\renewcommand{\thesection}{S.\arabic{section}}
\renewcommand{\theequation}{S.\arabic{equation}}
\renewcommand{\thealgocf}{S.\arabic{algocf}}

\section*{Supplementary materials}

\section{Generalization of explained variance}\label{supp:sec:exp_deviance}

In the Gaussian framework, we assume that $X_{ij} \sim \mathcal{N}(\mu_{ij}, 1)$ since data are preliminary centered and scaled in PCA. Under the assumptions of independence between observations, the log-likelihood is then in matrix notation:
\[
\begin{aligned}
\log p(\mb X\,\vert\,\mb M) && = & \sum_{i=1}^n\sum_{j=1}^m \log p(x_{ij}\,\vert\,\mu_{ij}) \\
&& = & \sum_{i=1}^n\sum_{j=1}^m (x_{ij} -  \mu_{ij})^2\\
&& = & \Vert \mb X - \mb M \Vert_F^2
\end{aligned}
\]
where $\mb M = [\mu_{ij}]$ is the matrix of Gaussian expectation and $\Vert \cdot \Vert_F^2$ is the squared Frobenius norm. In the generalized PCA framework \citep{collins2001}, we are looking for $\mb U \in \mathbb{R}^{n \times K}$ and $\mb V \in \mathbb{R}^{m \times K}$ such that $\mb M = \mb U\tr{\mb V}$. Thanks to \cite{eckart1936} theorem, best $\mb U$ and $\mb V$ minimizing the Frobenius norm between $\mb X$ and $\mb U\tr{\mb V}$ are given by Singular Value Decomposition (SVD) of $\mb X$, and optimal $\mb U$ exactly corresponds to the principal components from the PCA, which highlights the link between PCA, SVD and Gaussian framework.

In this Gaussian framework, the explained deviance defined in \cref{eq:exp_deviance} can be rewritten as
\[
\begin{aligned}
\%\text{dev}&& = & \frac{\log p(\mb X\,\vert\,\mb M = \hat{\mb U}\tr{\hat{\mb V}}) - \log p(\mb X\,\vert\,\mb M = \II_n \bar{\mb X})}{\log p(\mb X\,\vert\,\mb M = \mb X) - \log p(\mb X\,\vert\,\mb M = \II_n \bar{\mb X})},
\end{aligned}
\]
since the saturated model corresponds to $\mb M = \mb X$ in this case. It follows that
\[
\begin{aligned}
\%\text{dev}&& = & \frac{\Vert \mb X - \hat{\mb U}\tr{\hat{\mb V}} \Vert_F^2 - \Vert \mb X - \II_n \bar{\mb X} \Vert_F^2}{\Vert \mb X - \mb X \Vert_F^2 - \Vert \mb X - \II_n \bar{\mb X} \Vert_F^2}.
\end{aligned}
\]
In addition, we have that $ \bar{\mb X} = 0$ thanks to the pre-centering, and the formulation becomes:
\[
\begin{aligned}
\%\text{dev}&& = &1 -  \frac{\Vert \mb X - \hat{\mb U}\tr{\hat{\mb V}} \Vert_F^2}{\Vert \mb X \Vert_F^2}, \\
&& = & 1 - \frac{\sum_{k=K+1}^{\text{rk}(\mb X)} \sigma_k^2}{\sum_{\ell=1}^{\text{rk}(\mb X)} \sigma_{\ell}^2} \\
&& = & \frac{\sum_{k=1}^{K} \sigma_k^2}{\sum_{\ell=1}^{\text{rk}(\mb X)} \sigma_{\ell}^2},
\end{aligned}
\]
where $\text{rk}(\mb X)$ is the rank of $\mb X$ and $\sigma_1>\hdots>\sigma_{\text{rk}(\mb X)}$ the singular values of $\mb X$ (given by the SVD). The criterion corresponds exactly to the percentage of explained variance computed in PCA. Thus our percentage of explained deviance can be viewed as a generalization of this criterion to other distributions in the exponential family.

\section{Identifiability issues}\label{supp:sec:identifiability}

\subsection{Factor order}\label{supp:sec:order}

Gamma-Poisson factor model suffers from an identifiability issue regarding the order of factors. Unlike PCA, the components of model-based factor models are not orthogonal and can not be ordered naturally since the associated likelihood is identifiable up to a permutation of factors. Thus we propose an ordering defined by the cumulative Bregman divergence:
$k \mapsto D\big(\mb X\,\vert\,\hat{\mb U}_{1:k}\tr{(\hat{\mb V}_{1:k})}\big)$. In addition, we mention that the different GaP factor models are not nested when the dimension $K$ increases (as in the NMF), thus the factor estimates should be all computed for every choice of dimension $K$, contrary to PCA.


sub\section{Scaling effect in GaP factor model}\label{supp:sec:scaling}

As stated in \cite{dikmen2012}, GaP factor models suffer from identifiability issues, due to the scaling of the Gamma prior parameters $\mbg\alpha$ and $\mbg\beta$. Indeed, considering $\alpha_{k,2}^* = \eta_k\,\alpha_{k,2}$ and $\beta_{k,2}^* = (\eta_k)^{-1}\,\beta_{k,2}$ for fixed values $\eta_k$, and using the scaling property of the  Gamma distribution: if $U_{ik} \sim \text{Gamma}(\alpha_{k,1}, \alpha_{k,2})$ then $\eta_k\, U_{ik} \sim \Gam(\alpha_{k,1}, \eta_k^{-1}\alpha_{k,2})$. We show  (c.f. below) that the joint log-likelihood regarding $\mb U\mb H^{-1}$ and $\mb V\mb H$ with $\mb H = \text{diag}(\eta_k)_{k=1:K}$ verifies:
\begin{equation}
	\begin{aligned}
	&&&	\log p(\mb X, \mb U\mb H^{-1}, \mb V\mb H\,|\,\mbg\alpha_1, \mb H \mbg\alpha_2, \mbg\beta_1, \mb H^{-1}\mbg\beta_2)\\
		& = && \log p(\mb X, \mb U, \mb V\,|\,\mbg \alpha_1, \mbg \alpha_2, \mbg\beta_1, \mbg\beta_2) + (n-p) \sum_k \log(\eta_k)
	\end{aligned}
	\label{supp:eq:scaling}
\end{equation}
When $n=p$, there is an identifiability issue regarding the scaling of the parameters $\alpha_{k,2}$ and $\beta_{k,2}$, because different values lead to the same joint log-likelihood. In such case, a solution will be to fix the scale parameters $\alpha_{k,2}$ and $\beta_{k,2}$ to avoid the scaling effect. When $n\ne p$, the only problem is a potential solution with infinite norm with $\alpha_{k,2}\to 0$ and $\beta_{k,2} \to \infty$ or vice-versa \citep[c.f.][]{dikmen2012}. When considering zero-inflation or sparsity in the model, \cref{supp:eq:scaling} holds regarding the parameters of the Gamma prior distributions and we have to consider the same precaution. However, in practice we did not encounter such sequence of diverging parameters.

\paragraph{Proof of \cref{supp:eq:scaling}.} We set, $\alpha_{k,2}^* = \eta_k\,\alpha_{k,2}$ and $\beta_{k,2}^* = (\eta_k)^{-1}\,\beta_{k,2}$ for fixed values $\eta_k>0$. We use the scaling property of the Gamma distribution: if $U_{ik} \sim \text{Gamma}(\alpha_{k,1}, \alpha_{k,2})$ then $\eta_k\, U_{ik} \sim \Gam(\alpha_{k,1}, (\eta_k)^{-1}\alpha_{k,2})$. The joint log-likelihood regarding $\mb U\mb H^{-1}$ and $\mb V\mb H$ with $\mb H = \text{diag}(\eta_k)_{k=1:K}$ is then:

\[
\begin{aligned}
&&& \hspace{-1cm} \log p(\mb X, \mb U\mb H^{-1}, \mb V\mb H\,|\,\mbg\alpha_1, \mb H \mbg\alpha_2, \mbg\beta_1, \mb H^{-1}\mbg\beta_2)\\
& = && \sum_{i,j,k} \log p\big(x_{ij}\,\vert\,\{(\eta_k)^{-1}\,u_{ik}, \eta_k\,v_{jk}\}_{k=1:K}\big)\\
&&& + \sum_{i,k} \log p\big(\eta_k^{-1}\,u_{ik}\,;\,\alpha_{k,1}, \eta_k\,\alpha_{k,2}\big) \\
&&& + \sum_{j,k} \log p\big(\eta_k\, v_{jk}\,;\,\beta_{k,1}, (\eta_k)^{-1}\,\beta_{k,2}\big) \\
& = && \log p(\mb X\,|\, \mb U, \mb V) + \log p(\mb U\,;\,\mbg \alpha_1, \mbg \alpha_2) + \sum_{i=1}^n \sum_{k=1}^K \log(\eta_k)\\
&&& + \log p(\mb V\,;\, \mbg\beta_1, \mbg\beta_2) + \sum_{j=1}^p \sum_{k=1}^K -  \log(\eta_k)\\
& = && \log p(\mb X, \mb U, \mb V\,|\,\mbg \alpha_1, \mbg \alpha_2, \mbg\beta_1, \mbg\beta_2) + (n-p) \sum_k \log(\eta_k)
\end{aligned}
\]

\section{Variational inference algorithm}

\cref{supp_mat:fig:varinf} describes the variational framework (for the GaP factor model) that we extended to develop our approach.

\begin{figure}
\centering
{
\begin{tabular}{ccccc}
\begin{tabular}{l}
\textbf{The model}$^{(*)}$ \\
$X_{ij} = \sum_k Z_{ijk}$ \\
$Z_{ijk}\,|\,U_{ik}, V_{jk} \sim \Poi(U_{ik}\,V_{jk})$ \\
$U_{ik} \sim \Gam(\alpha_{k,1},\alpha_{k,2})$ \\
$V_{jk} \sim \Gam(\beta_{k,1},\beta_{k,2})$ \\
\end{tabular} & $\longrightarrow$ &
\begin{tabular}{c}
\textbf{Intractable} \\ \textbf{posterior}\\
\end{tabular} & $\longrightarrow$ & 
\begin{tabular}{|c|}
\hline
\textbf{Variational} \\ \textbf{framework} \\
\hline
\end{tabular} \\
&&&& $\Big\downarrow$ \\
\\
&& \begin{tabular}{|c|}
\hline
\textbf{Optimization} \\ of $J(q)$ \\
\hline
\end{tabular} & $\longleftarrow$ & \begin{tabular}{c}
\textbf{Approximate} \\ the \textbf{posterior}\\ by the distrib. $q$ \\
\end{tabular}
\\
&& $\swarrow$ \hspace{50pt} $\searrow$ \\
\\
\multicolumn{5}{c}{
\hspace{50pt} \begin{tabular}{cc}
\begin{tabular}{l}
\textbf{Variational distribution} \\
$U_{ik} \overset{q}\sim \Gam(a_{ik,1},a_{ik,2})$ \\
$V_{jk} \overset{q}\sim \Gam(b_{jk,1},b_{jk,2})$ \\
$(Z_{ijk})_k \overset{q}\sim \multin\Big(X_{ij}, (r_{ijk})_k\Big)$ \\
\end{tabular} \hspace{3pt} & \hspace{3pt} \begin{tabular}{l}
\textbf{Complete conditional} \\
$U_{ik}\,|\,\cond\ \sim \Gam\Big(\mbg\eta_{ik}(\cond)\Big)$ \\
$V_{jk}\,|\,\cond\ \sim \Gam\Big(\mbg\eta_{jk}(\cond)\Big)$ \\
$(Z_{ijk})_k\,\vert\,\cond\ \sim \mathcal{M}\Big(X_{ij}, (\rho_{ijk})_k\Big)$ \\
\end{tabular} \\
\end{tabular}
} \\
\\
&& $\searrow$ \hspace{50pt} $\swarrow$ \\
\\
&& \begin{tabular}{|c|}
\hline
\textbf{Inference of $q$} \\
\hline
\end{tabular} \\
\\
\multicolumn{5}{l}{
{\scriptsize $(*)$ with conditional independence between the $Z_{ijk}$'s and independence between the $U_{ik}$'s and $V_{jk}$'s}}\\
\\
\end{tabular}}
\caption{Variational inference to approximate the posterior of the model, based on the optimization of the ELBO that required to derive the full conditional. The notation $\overset{q}\sim$ refers to the variational distribution.}
\label{supp_mat:fig:varinf}
\end{figure}

\subsection{Full conditional distributions}\label{supp:sec:meth:conditional}

In our factor model all full conditionals are tractable. Thanks to the Gamma-Poisson conjugacy, the full conditionals of $U_{ik}$ and $V'_{jk}$ are Gamma distributions. The proof is based on the Bayes rule and the distribution of the latent variables $\mb Z$, that are actually necessary to derive $p(U_{ik}\,\vert\,\cond\,)$ and $p(V'_{jk}\,\vert\,\cond\,)$. The full conditional of the vector $ \mb Z_{ij}$ is also explicit, being a Multinomial distribution \citep{zhou2012} when $D_{ij} \ne 0$ and deterministic null when $D_{ij}=0$, i.e. $(Z_{ijk})_k\,\vert\,\cond\ \sim D_{ij}\, \mathcal{M}\big(X_{ij}, (\rho_{ijk})_k\big)$. Here the Multinomial probabilities $(\rho_{ijk})_k$ depend on $(S_{jk}, U_{ik}, V'_{jk})_k$, and quantify the prior contribution of factor $k$ to the observations $X_{ij}$, i.e. $$\rho_{ijk} = \frac{S_{jk}\,U_{ik}\,V'_{jk}}{\sum_{\ell}\,S_{j\ell}\, U_{i\ell}\,V'_{j\ell}}.$$ 
This point justifies why the variational distribution is based on the vector $\mb Z_{ij}$ instead of taking each $Z_{ijk}$ separately. Note that if the $S_{jk}$ are null for all $k$ or if $D_{ij} = 0$ (i.e. $X_{ij}=0$), the vector $(Z_{ijk})_k$ is deterministic and takes null values. We summarize the full conditionals in the sparse ZI-GaP factor model regarding $U_{ik}$, $V_{jk}$ and $(Z_{ijk})_k$, that are defined such as:
\begin{equation}
\begin{aligned}
& U_{ik}\,|\,\cond\ \sim \Gam(\alpha_{k,1} + \msum_j\, D_{ij}\,S_{jk}\,Z_{ijk},\ \alpha_{k,2} + \msum_j\, D_{ij}\,S_{jk}\,V'_{jk}) \, , \\
& V'_{jk}\,|\,\cond\ \sim \Gam(\beta_{k,1} + \msum_i\, D_{ij}\,S_{jk}\,Z_{ijk},\ \beta_{k,2} + \msum_i\, D_{ij}\,S_{jk}\,U_{ik}) \, , \\
& (Z_{ijk})_k\,\vert\,\cond\ \sim D_{ij}\,\mathcal{M}\Big(X_{ij}, (\rho_{ijk})_k\Big) \, , \\
\end{aligned}
\label{eq:comp_cond}
\end{equation}

\paragraph{Zero Inflation.} Regarding the zero-inflation indicators, $D_{ij}$ is a binary variable, its distribution is either deterministic or Bernoulli. When the entry $X_{ij}$ is non null, $D_{ij}$ is certainly equal to one. When $X_{ij} = 0$, the full conditional is explicit and the Bernoulli probability only depends on the prior over $D_{ij}$ and the probability that $X_{ij}$ is null. It can be formulated as follows:
\[
p(D_{ij} = 1\,|\,\cond\,)  \propto  \pi_{j}^D\ e^{-\sum_k\,S_{jk}\,U_{ik}\,V'_{jk}}\, .
\]

\paragraph{Sparsity and variable selection.} The sparsity indicator $S_{jk}$ is also a binary variable and its full conditional is also an explicit Bernoulli distribution. It depends on the prior over $S_{jk}$ and the probability that gene $j$ contributes to the components $k$, quantified by the joint distribution on $(Z_{ijk})_i$, thus:
{\small \[
p(S_{jk} = 1\,\vert\,\cond\,) \propto \pi_{j}^\text{s} \times \mprod_i\, \exp(-S_{jk}\,U_{ik}\,V'_{jk})\ (S_{jk}\,U_{ik}\,V'_{jk})^{Z_{ijk}}.
\]}

\subsection{Derivation of variational parameters}\label{supp:sec:meth:var_update}

\paragraph{Variational parameters of factors.} We derive the stationary point formulation for the variational parameters regarding $U_{ik}$ and $V'_{jk}$, being explicitly (directly derived from the partial derivatives of $J(q)$):
\[
\begin{aligned}
& \mb a_{ik} = \Big( \alpha_{k,1} + \msum_j\, \hat{D}_{ij}\, \hat{S}_{jk}\,\hat{Z}_{ijk}\,,\ \alpha_{k,2} + \msum_j\, \hat{D}_{ij}\, \hat{S}_{jk}\,\hat{V'}_{jk} \Big)^T\\
& \mb b_{jk} = \Big( \beta_{k,1} + \hat{S}_{jk}\,\msum_i\, \hat{D}_{ij}\, \hat{Z}_{ijk}\,,\ \beta_{k,2} + \hat{S}_{jk}\,\msum_i\,\hat{D}_{ij}\, \hat{U}_{ik}\Big)^T,
\end{aligned}
\]
which generalizes formulations from standard GaP factor model \citep{cemgil2009}. As for variable $Z_{ijk}$, its posterior distribution depends on parameter $r_{ijk}$ with the relation $\log(r_{ijk})=\EE_q [\log (\rho_{ijk})]$. Hence, the variational distribution on $(Z_{ijk})_k$ naturally depends on the selection indicator $S_{jk}$  (since our model focuses on loadings selection). In particular, the variational parameter $r_{ijk}$ depends on $S_{jk}$, through a specific term $\EE_q[\log(S_{jk}\,V'_{jk})]$ that is computed using the variational distribution of $S_{jk}$ (a Bernoulli distribution of parameter $p_{jk}^S$). To proceed, we introduce $\tilde{S}_{jk}$, the discretized predictor of $S_{jk}$ such that $\tilde{S}_{jk} = \II_{\{p_{jk}^S > \tau\} }$, where $\tau$ is a threshold specified by the user (for instance $0.5$). Then, the formulation of the optimal variational parameter $r_{ijk}$ is approximated by:
\[
\begin{aligned}
&  r_{ijk} = \frac{\tilde{S}_{jk}\,\exp\Big( \hat{\log U}_{ik} + \hat{\log V'}_{jk}\Big)}{\sum_{\ell} \tilde{S}_{j\ell}\, \exp\Big(\hat{\log U}_{i\ell} + \hat{\log V'}_{j\ell}\Big)}.
\end{aligned}
\]

\paragraph{Variational dropout proportion.} Regarding the zero-inflated probabilities $p_{ij}^D$, when $X_{ij}\neq 0$, the posterior is explicit since $D_{ij}=1$ with probability one. Hence, only the case $X_{ij} = 0$ requires a variational inference. As stated previously, the full conditional is explicit and it is possible to derive and optimize the ELBO (based on the natural parametrization of the Bernoulli distribution in the exponential family). Eventually, $p_{ij}^D$ is computed as:
\[
\begin{aligned}
\begin{aligned}
\text{logit}(p_{ij}^{D})
& = && \text{logit}(\pi_{j}^D) - \msum_k\,\hat{S}_{jk}\, \hat{U}_{ik} \hat{V'}_{jk}\, ,
\end{aligned}
\end{aligned}
\]
where the Bernoulli prior probability $\pi_{j}^D$ is corrected by $\EE_q[\log\PP(X_{ij}=0)]$ to account for the probability of $X_{ij}$ being a true zero.

\paragraph{Variational Selection probability.} Concerning the sparse indicator $S_{jk}$, the natural parametrization of the Bernoulli distribution is based on the logit of the Bernoulli probability. Hence we can write an explicit formulation of the ELBO regarding $p_{jk}^S$ based on the full conditional on $S_{jk}$. Following this formulation, the stationary point  $p_{jk}^S$ verifies:
\[
\begin{aligned}
\text{logit}(p_{jk}^S)
= \text{logit}(\pi_{j}^S) - \msum_i & \hat{D}_{ij}\, \hat{U}_{ik}\, \hat{V'}_{jk} \\ & + \hat{D}_{ij}\, \hat{Z}_{ijk}\,\big(\hat{\log U}_{ik} + \hat{\log V'}_{jk} \big)\, .
\end{aligned}
\]
This corresponds to a correction of the Bernoulli prior probability $\pi_{j}^S$, depending on the quantification of the contribution of gene $j$ to component $k$ in all individuals, i.e. $\EE_q[\msum_i \log p(Z_{ijk})]$.

\subsection{Derivation of prior parameters}\label{supp:sec:meth:prior_update}

The hyper-parameters of priror distribution regarding $U_{ik}$, $V_{jk}$, $D_{ij}$ and $S_{jk}$ are updated within the M-step such that (respectively):
{\small \[
\begin{aligned}
& \alpha_{k,1} &=& \psi^{-1} \left(\log \alpha_{k,2} + \frac{1}{n}\sum_i \hat{\log U}_{ij} \right), 
& \alpha_{k,2} &=&\frac{\alpha_{k,1}}{\sum_i \hat{U}_{ij}/n}, \\
& \beta_{k,1} &=& \psi^{-1} \left(\log \beta_{k,2} + \frac{1}{p}\sum_j \hat{\log V'}_{ij} \right),
& \beta_{k,2} &=&\frac{\beta_{k,1}}{\sum_j \hat{V'}_{ij}/p}, \\
& \pi_{j}^{\text{D}} &=& \frac{1}{n}\sum_i p_{ij}^{\text{D}},
& \pi_{j}^{\text{S}} &=& \frac{1}{K}\sum_k p_{jk}^{\text{S}}, \\
\end{aligned}
\]}
where $\psi$ is the digamma function, i.e. the derivative of the log-Gamma function. Its inverse is computed thanks to the method proposed in \citet[][appendix C]{minka2000}. Recalling that, for a variable $U\sim\Gam(\alpha_1,\alpha_2)$, $\EE[U] = \alpha_1/\alpha_2$ and $\EE[\log U] = \psi(\alpha_1) -  \log \alpha_2$, the update rule for the Gamma prior parameters on $U_{ik}$ corresponds to averaging the moments and log-moments of the variational distribution on $U_{ik}$ over $i$ (similarly for $V_{jk}$ over $j$). Regarding the Bernoulli prior parameters $\pi_{j}^{\text{D}}$, the update rule is also an average of the corresponding variational parameter over $i$ (similarly for $\pi_{j}^{\text{S}}$ over $k$).

\subsection{Algorithm}

Our pCMF algorithm is summarized in \Cref{supp:algo:varEM}. In the initialization step, each variational Gamma shape parameter $a_{ik,1}$ and $b_{jk,2}$ are sampled from a Gamma distribution \citep{zhou2012}.  Each variational Gamma rate parameter $a_{ik,2}$ and $b_{jk,2}$ are set to 1 (to avoid scaling effect between shape and rate parameters). Each variational dropout probability $p_{ij}^D$ is initialized with the corresponding indicator $\delta_0(X_{ij})$. Each variational sparsity probability $p_{jk}^S$ is initialized with the corresponding with the threshold value $\tau\in(0,1)$. In addition, all prior hyper-parameters are initialized following update rules based on variational parameters (defined in \cref{sec:meth:prior_update} in the paper).\\

The convergence is assessed by computing the normalized gap between two successive parameter values across iterations. When the updates does not modify the values of the parameters, we can consider that we reach a fixed point and thus the optimum. In addition, to overcome potential issue related to local optimum, the algorithm is run several times with different random initialization and the best seed (regarding the ELBO criterion) is kept.

\RestyleAlgo{boxruled}
\begin{algorithm}
\KwData{count matrix \mb X}
\KwResult{factors $\hat{\mb U}$ and $\hat{\mb V}$}
\KwInit{random initialization of variational parameters, prior hyper-parameters are updated accordingly}
\While{No convergence}{
Update variational parameters (see \cref{sec:meth:var_update} in the paper)\;
Update prior pararmeters (see \cref{sec:meth:prior_update} in the paper)\;
}
\caption{Variational EM algorithm}
\label{supp:algo:varEM}
\end{algorithm}

\section{Data generation}\label{supp:sec:data_gen}

We set the hyper-parameters $(\alpha_{k,1},\alpha_{k,2})_k$ and $(\beta_{k,1},\beta_{k,2})_k$ of the Gamma prior distributions on $U_{ik}$ and $V_{jk}$ to generate structure in the data, i.e. groups of individuals and groups of variables.

\paragraph{Generation of $\mb U$.} In practice, individuals $i=1,\hdots,n$ are partitioned into $N$ balanced groups, denoted by $\mathcal{U}_1,\hdots,\mathcal{U}_N$. To do so, we generate a matrix $\mb U$ with blocks on the diagonal. Each block, denoted by $\mathcal{B}_{\mb U,g}$ contains $n/N$ rows and $K/N$ columns.
Each entry $U_{ik}$ in each block $\mathcal{B}_{\mb U,g}$ ($g=1,\hdots,N$) is drawn from a Gamma distribution $\Gam(1,1/\alpha_g)$ with a rate parameter depending on $\alpha_g>0$ (different for each group). All entries $U_{ik}$ that are not in the diagonal blocks of $\mb U$ are drawn from a Gamma distribution $\Gam(1, 1/((1-\theta_u) \bar{\alpha})$ where $\bar{\alpha}$ is the average of the $\alpha_g$'s across $g$, and $\theta_u \in(0,1)$ quantifies how much the groups of individuals are distinct. Hence, each groups of individuals $\mathcal{U}_g$ corresponds to a block $\mathcal{B}_{\mb U,g}$. Thus, this generation pattern requires that $K>N$. In practice, we fix $\alpha_g \in\{100,250\}$, we use $\theta = 0.5$ or $0.8$ (for low or high separation respectively) and $N=3$ groups of individuals.

\paragraph{Generation of $\mb V$.}

The question of simulating data based on a sparse representation $\mb V$ of the variables in our context of matrix factorization is not straightforward. Indeed, if we impose that some variables $j$ do not contribute to any component $k$, i.e. that $V_{jk}$ is null for any $k$, then $\msum_k\,U_{ik}\,V_{jk}$ is always null for $i=1,\hdots,n$. Thus, the recorded data entry $X_{ij}$ will be deterministic and null for any observation $i$ (i.e. the $j^\text{th}$ column in $\mb X$ will be null). There is no interest to generate full columns of null values in the matrix $\mb X$, since it is unnecessary to use a statistical analysis to determine that a column of zeros will not be informative. This question is not an issue about the formulation of the model, but rather concerns the generation of non informative columns in $\mb X$ that will correspond to null rows in the matrix $\mb V$.\\

To overcome this issue, we use the following generative process. The variables $j=1,\hdots,p$ are first partitioned into two groups $\mathcal{V}_0$ and $\mathcal{V}_{\varnothing}$ of respective sizes $m_0$ and $m-m_0$ (with $m_0\leq m$). The $m_0$ variables in $\mathcal{V}_0$ will represent the pertinent variables for the lower dimensional representation, whereas variables in $\mathcal{V}_{\varnothing}$ will be considered irrelevant or noise. The matrix $\mb V$ will be a concatenation of two matrices $\mb V^0$ and $\mb V^{\varnothing}$:
\[
\mb V_{m\times K} = \left( \begin{array}{c} \mb V^0 \\ \mb V^{\varnothing} \end{array}\right)
\]
The ratio $m_0/m$ sets the expected degree of sparsity in the model. In practice, we generate $m_0/m$ from a Beta distribution, so that in average $m_0/m \in \{0.2, 0.4, 0.6, 0.8\}$ corresponding to different proportions of noisy genes (between $20$ and $80\%$ of noisy genes). \\

To simulate dependency between recorded variables, we generate groups of variables in the set $\mathcal{V}_0$ of pertinent variables. We use a similar strategy as the one used to simulate $\mb U$. $\mathcal{V}_0$ is partitioned into  $M$ balanced groups, denoted by $\mathcal{V}_1,\hdots,\mathcal{V}_M$. We generate the corresponding matrix $\mb V^0$ with blocks on the diagonal. Each block, denoted by $\mathcal{B}_{\mb V,g}$ contains $m_0/M$ rows and $K/M$ columns:
Each entry $V_{jk}$ in each block $\mathcal{B}_{\mb V,g}$ ($g=1,\hdots,M$) is drawn from a Gamma distribution $\Gam(1, 1/\beta)$ with a rate parameter depending on $\beta>0$. All entries $V_{jk}$ that are not in the blocks on diagonal are drawn from a Gamma distribution $\Gam(1, 1/((1-\theta_v) \beta)$, where $\theta_v \in(0,1)$ quantifies how much the groups of individuals are distinct. Hence, each groups of individuals $\mathcal{V}_g$ corresponds to a block $\mathcal{B}_{\mb V,g}$. Again, this generation pattern requires that $K>M$. In practice, we fix $\beta =80$, we use $\theta_v = 0.8$ and $M=2$ groups of variables.

In addition, all $V_{jk}$ in $\mb V^{\varnothing}$ (noisy genes) are drawn from a Gamma distribution $\Gam(1, 1/((1-\theta_v) \beta)$, so that $\EE[V_{jk}]$ will not be structured according to groups.

\paragraph{Generation of $\mb X$.} The data are simulated according to their conditional Poisson distribution in the model i.e. $\Poi(\msum_k\,u_{ik}\,v_{jk})$. In practice, we want to consider zero-inflation in the model, thus we consider the Dirac-Poisson mixture and simulate $X_{ij}$ according to the following conditional distribution:
\[
X_{ij}\,|\,(U_{ik}, V_{jk})_k, D_{ij} \sim (1- D_{ij}) \times \delta_0 + D_{ij} \times \Poi(\msum_k\,U_{ik}\,V_{jk})\, ,
\]
where the dropout indicator $D_{ij}$ is drawn from a Bernoulli distribution $\mathcal{B}(\pi_j^D)$, the proportion of dropout events is set by the probability $\pi_j^D$. To generate data without dropout events, we just have to set $D_{ij} = 1$ for any couple $(i,j)$, i.e. $\pi_j^D=1$ for any $j$.\\

In practice, we fix $K=40$, $n=100$ and $m=800$ to simulate our data. We generate different level of zero-inflation: $\pi_j^D$ is drawn from a beta distribution so that in average it lies in $\{0.3, 0.5, 0.7, 0.9\}$.

\section{Softwares}\label{supp:sec:soft}

The Poisson-NMF is from the \texttt{NMF} R-package \citep{gaujoux2010}, ZIFA from the \texttt{ZIFA} Python-package \citep{pierson2015}, the sparse PCA from the \texttt{PMA} R-package \citep{witten2009} and t-SNE from the \texttt{Rtsne} R-package \citep{krijthe2015}. Computation of adjusted Rand Index was done thanks to the \texttt{mclust} R-package \citep{fraley2002}.

\FloatBarrier

\newpage

\section{Additional results}\label{supp:sec:simu:results}

\subsection{Selection accuracy}

\begin{figure}[!tpb]
\centering
\begin{subfigure}{.38\linewidth}
\includegraphics[width=.95\linewidth]{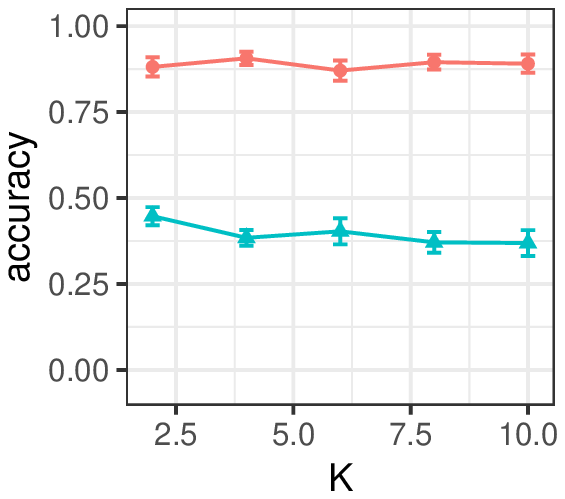}
\caption{}
\label{supp:fig:simu:quant:3groups:accsel:ncomp}
\end{subfigure}
\begin{subfigure}{0.6\linewidth}
\includegraphics[width=.94\linewidth]{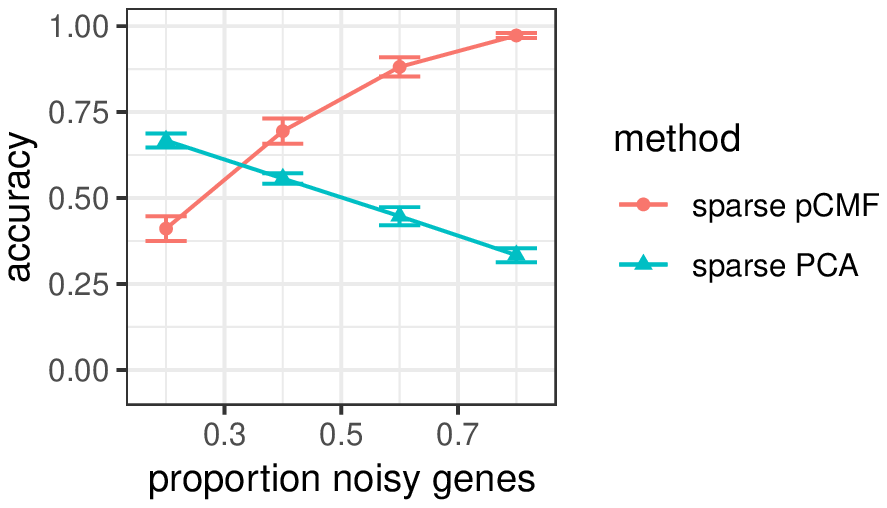}
\caption{}
\label{supp:fig:simu:quant:3groups:accsel:noise}
\end{subfigure}
\caption{Selection accuracy depending on the dimension $K$ (\ref{supp:fig:simu:quant:3groups:accsel:ncomp}) with a proportion of noisy genes set to $60\%$ and the proportion of noisy genes (\ref{supp:fig:simu:quant:3groups:accsel:noise}) with $K$ set to 2. Average values and deviation are estimated across 50 repetitions.}
\label{supp:fig:simu:quant:3groups:accsel}
\end{figure}

See \cref{supp:fig:simu:quant:3groups:accsel}.

\subsection{Clustering}

\begin{figure}[!tpb]
\centering
\begin{subfigure}{.95\linewidth}
\includegraphics[width=.95\linewidth]{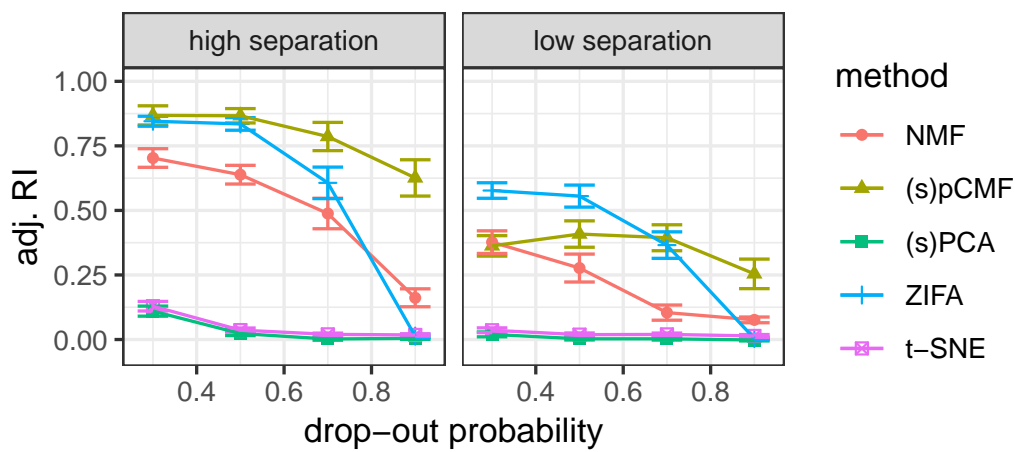}
\caption{}
\label{supp:fig:simu:quant:3groups:adjri:U}
\end{subfigure}
\begin{subfigure}{0.95\linewidth}
\includegraphics[width=.95\linewidth]{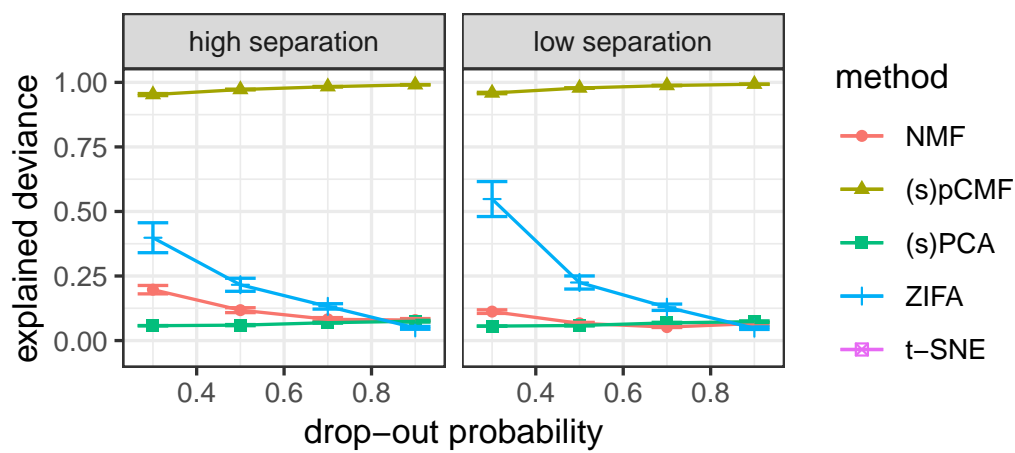}
\caption{}
\label{supp:fig:simu:quant:3groups:expdev:U}
\end{subfigure}
\caption{Adjusted Rand Index (\ref{supp:fig:simu:quant:3groups:adjri:U}) for the clustering on $\hat{\mb U}$ versus the true groups of cells; and explained deviance (\ref{supp:fig:simu:quant:3groups:expdev:U}) depending on the probability used to generate dropout events, for different levels of separability between cell groups. Average values and deviation are estimated across 50 repetitions.}
\label{supp:fig:simu:quant:3groups:U}
\end{figure}

\begin{figure}[!tpb]
\centering
\begin{subfigure}{.95\linewidth}
\includegraphics[width=.95\linewidth]{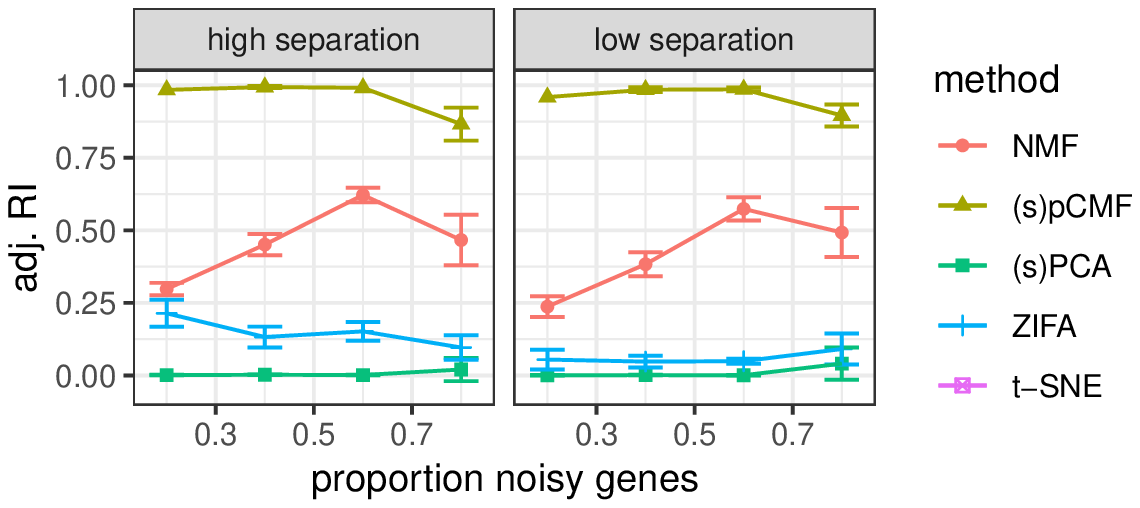}
\caption{}
\label{supp:fig:simu:quant:3groups:adjri:V}
\end{subfigure}
\begin{subfigure}{0.95\linewidth}
\includegraphics[width=.95\linewidth]{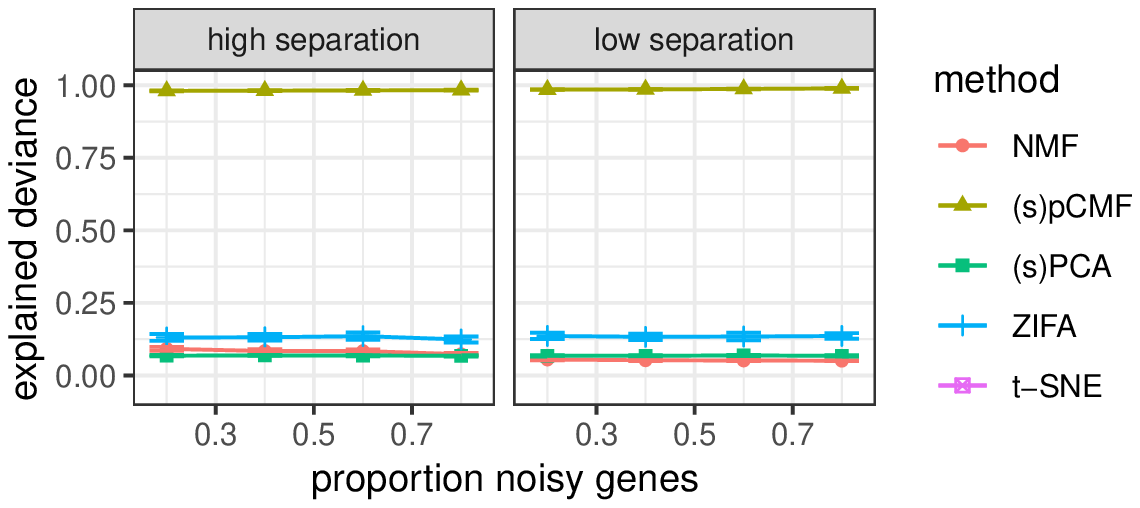}
\caption{}
\label{supp:fig:simu:quant:3groups:expdev:V}
\end{subfigure}
\caption{Adjusted Rand Index (\ref{supp:fig:simu:quant:3groups:adjri:V}) for the clustering on $\hat{\mb V}$ versus the true groups of genes; and explained deviance (\ref{supp:fig:simu:quant:3groups:expdev:V}) depending on the proportion of noisy genes, for different levels of separability between cell groups. Average values and deviation are estimated across 50 repetitions.}
\label{supp:fig:simu:quant:3groups:V}
\end{figure}

See \cref{supp:fig:simu:quant:3groups:U,supp:fig:simu:quant:3groups:V}. We present results on simulations with different degree of separation between the groups of individuals.

\subsection{Computation time}\label{supp:sec:simu:time}

\begin{figure}
\centering
\includegraphics[width=.9\linewidth]{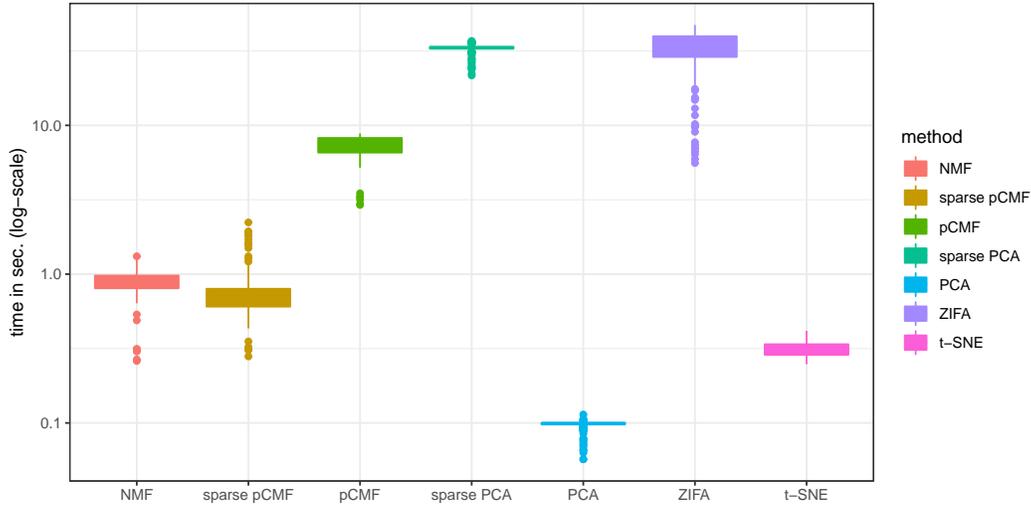}
\caption{Computation time on 8-CPU core for the different approaches, running time required to analyse simulated data with $n=100$ individuals and $m=800$ cells (50 repetitions).}
\label{supp:fig:simu:time}
\end{figure}

\cref{supp:fig:simu:time} shows average computation time for the different methods (pCMF, Poisson-NMF, SPCA, ZIFA) for a single run on a 8-core standard CPU with frequency between 2 and  2.5 GHz. All methods, including ours, have different levels of multi-threading and can benefit from multi-core CPU computations.\\

Our method sparse pCMF shows comparable computation time as state-of-the-art approaches as Poisson-NMF. The npn-sparse version pCMF is slower but still faster than ZIFA and sparse PCA (because the latter requires a cross-validation step to tune a penalty parameter). t-SNE is slightly faster but requires numerous run with different values for the perplexity parameter (here the timing corresponds to a run for a single perplexity value). The PCA is the gold standard regarding running time thanks to the efficiency of its algorithm based on the Singular Value Decomposition (SVD) algorithm. Packages from where the different methods can be found are detailed in \cref{supp:sec:soft}.\\

Eventually, we mention that our method is available in an R-package, however our algorithms are implemented in interfaced C++ for computational efficiency.

\subsection{Standard GaP versus our ZI sparse GaP factor model}\label{supp:sec:simu:gap}

\begin{figure}
\centering
\includegraphics[width=.9\linewidth]{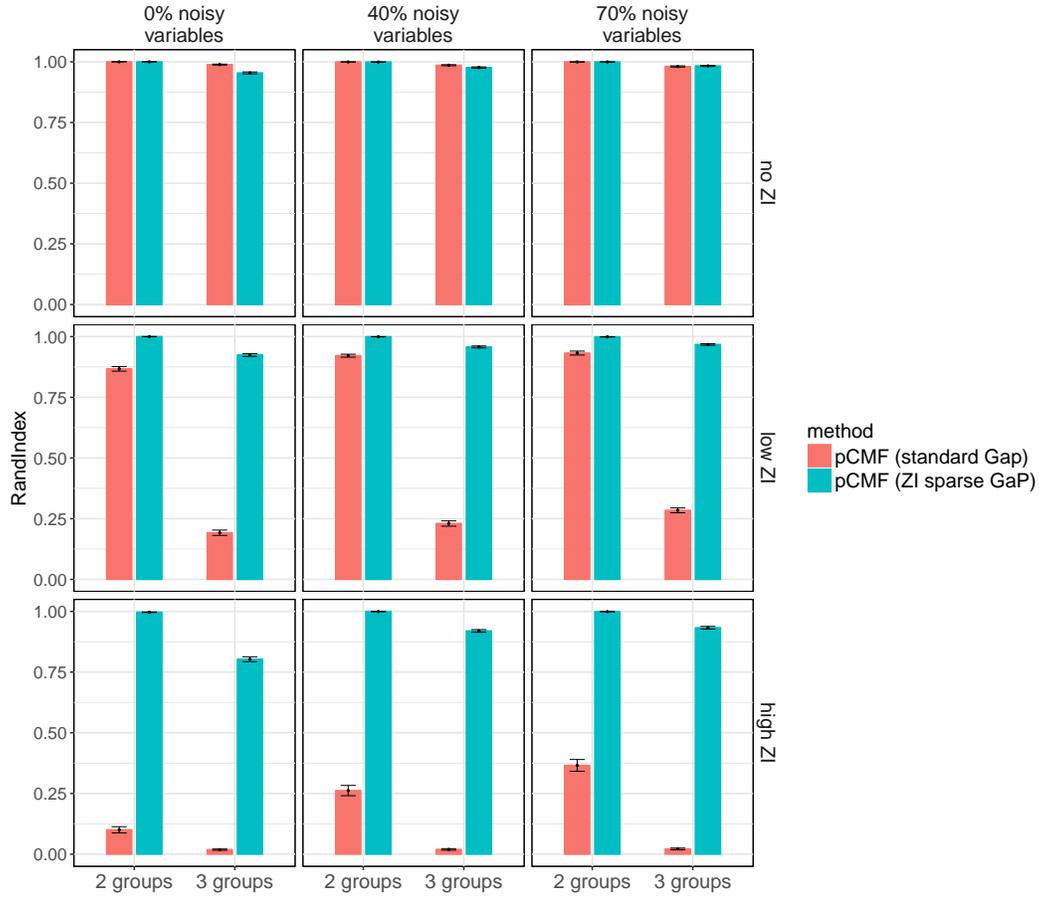}
\caption{Adjusted Rand Index comparing clusters found by a $\kappa$-means algorithm (applied to $\hat{\mb U}$ with $\kappa=2$) and the original groups of individuals, depending on the number of individual groups in the data, for different levels of zero-inflation and different proportion of noisy variables in the data. The number of components is set to $K=10$. Data are generated with $n=100$, $m=1000$. Average values and deviation are estimated across 100 repetitions.}
\label{supp:fig:simu:gap}
\end{figure}

\cref{supp:fig:simu:gap} illustrates the interest of our zero-inflated sparse Gamma-Poisson factor model compared to the standard Gamma-Poisson factor model, especially in presence of dropout events and noisy genes. Our method pCMF based on our ZI sparse GaP factor model performs as well as the pCMF based on the standard GaP factor model when there is no dropout events in the data, independently from the proportion of noisy genes. In addition, when the level of zero-inflation is higher, we can see that the ZI-specific model outperforms the standard ones, highlighting the interest of our approach.

\FloatBarrier

\newpage

\subsection{Additional scRNA-seq data analyses}\label{supp:sec:data}

The dataset from \cite{baron2016} is available here\footnote{\href{https://hemberg-lab.github.io/scRNA.seq.datasets/mouse/pancreas/}{https://hemberg-lab.github.io/scRNA.seq.datasets/mouse/pancreas/}}. The goldstandard and silverstandard datasets used in \cite{freytag2018} can be found here\footnote{\href{https://github.com/bahlolab/cluster\_benchmark\_data}{https://github.com/bahlolab/cluster\_benchmark\_data}} (we used the silverstandard dataset 5 which was the largest). The 3 previous datasets are stored based on the \texttt{SingleCellExperiment} R package \cite{lun2019}. The dataset from \cite{llorens-bobadilla2015} was available as supplementary data of their paper. They kindly shared with us the information about cell tags.

\subsubsection{\cite{llorens-bobadilla2015}}
We illustrate the performance of pCMF on a publicly available scRNA-seq datasets of neuronal stem cells \citep{llorens-bobadilla2015}. Neural stem cells (NSC) constitute an essential pool of adult cells for brain maintenance and repair. \cite{llorens-bobadilla2015} proposed a study to unravel the molecular heterogeneities of NSC populations based on scRNA-seq, and particularly focused on quiescent cells (qNSC). In their experiment, qNSC were transplanted in vivo in order to study their neurogenic activity. Following transplantation, 92 qNSC produced neuroblasts and olfactory neurons, whose transcriptome was compared with 21 astrocytes (CTX) and 27 transient amplifying progenitor cells (TAP). The authors used a PCA approach to reveal a continuum of ``activation state'', from astrocytes (low activation) to amplifying progenitor cells (TAP).

As stated before, we confront pCMF with other state-of-the-art approaches. The first visual result (c.f. \cref{supp:fig:data:visu:llorens:cells}) is that pCMF provides a slightly better representation of the continuum of activation described by \citeauthor{llorens-bobadilla2015} than PCA and t-SNE, which probably reflects a better modeling of the biological variations that exist between activation states. In practice, t-SNE was not able to highlight the different clusters of cells. The results from ZIFA are consistent with pCMF representation, which is a confirmation that the signal of this continuous activation state is strong in these data. These qualitative results are confirmed by clustering quantitative results (c.f. \cref{table:data:criteria} in the manuscript). The adjusted Rand Index computed after pCMF and ZIFA are similar and better than PCA.

\subsubsection{Representation of cells}

\begin{figure}[h]
\centering
\includegraphics[width=.99\linewidth]{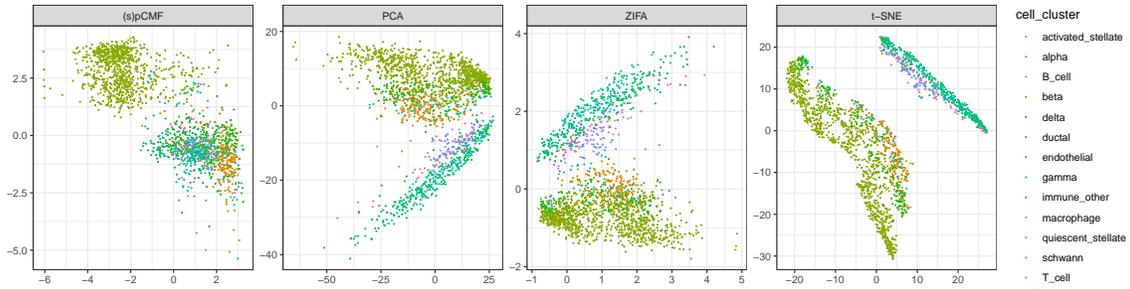}
\caption{Analysis of the scRNA-seq dataset from \cite{baron2016}, $1186$ cells, $6080$ genes. Visualization of the cells in a latent space of dimension 2.}
\label{supp:fig:data:visu:baron:cells}
\end{figure}

\begin{figure}[h]
\centering
\includegraphics[width=.99\linewidth]{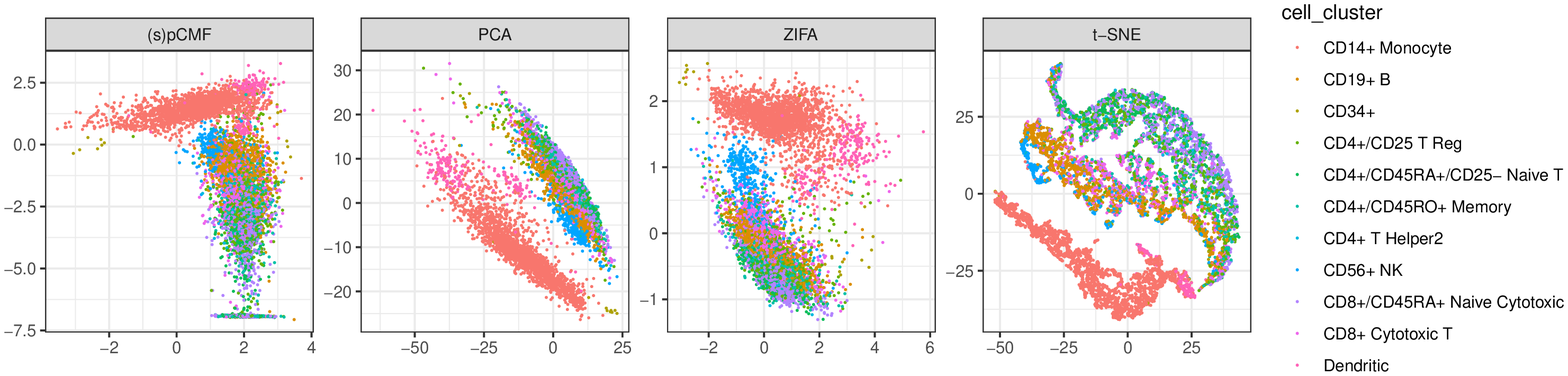}
\caption{Analysis of the silverstandard 5 scRNA-seq dataset from \cite{freytag2018}, $8352$ cells, $4547$ genes. Visualization of the cells in a latent space of dimension 2.}
\label{supp:fig:data:visu:silver:cells}
\end{figure}

\begin{figure}[h]
\centering
\includegraphics[width=.99\linewidth]{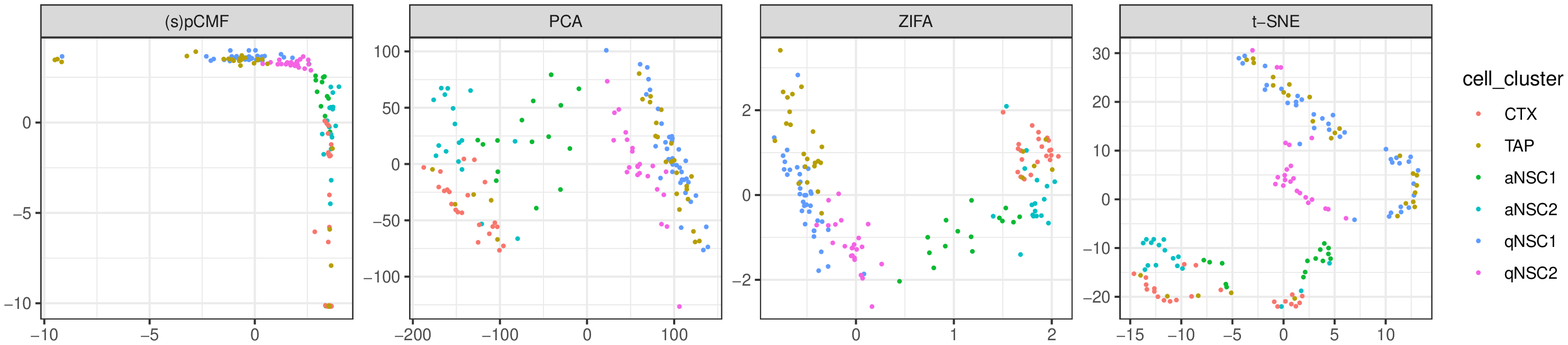}
\caption{Analysis of the scRNA-seq dataset from \cite{llorens-bobadilla2015}, $141$ cells, $13826$ genes. Visualization of the genes in a latent space of dimension 2.}
\label{supp:fig:data:visu:llorens:cells}
\end{figure}

See \cref{supp:fig:data:visu:baron:cells,supp:fig:data:visu:silver:cells,supp:fig:data:visu:llorens:cells}.

\subsubsection{Representation of genes}

\begin{figure}[h]
\centering
\includegraphics[width=.99\linewidth]{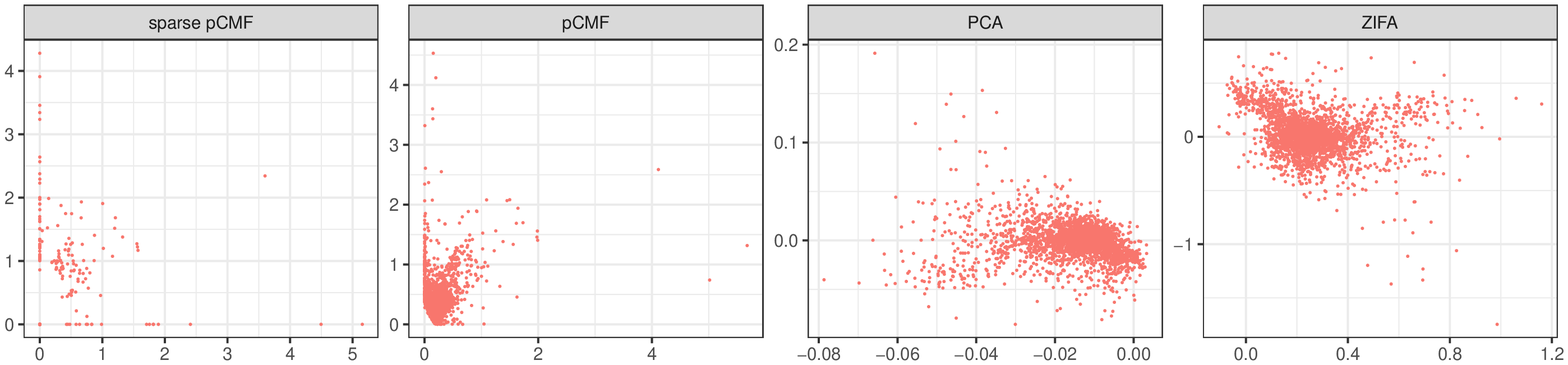}
\caption{Analysis of the scRNA-seq dataset from \cite{baron2016}, $1186$ cells, $6080$ genes. Visualization of the genes in a latent space of dimension 2.}
\label{supp:fig:data:visu:baron:genes}
\end{figure}

\begin{figure}[h]
\centering
\includegraphics[width=.99\linewidth]{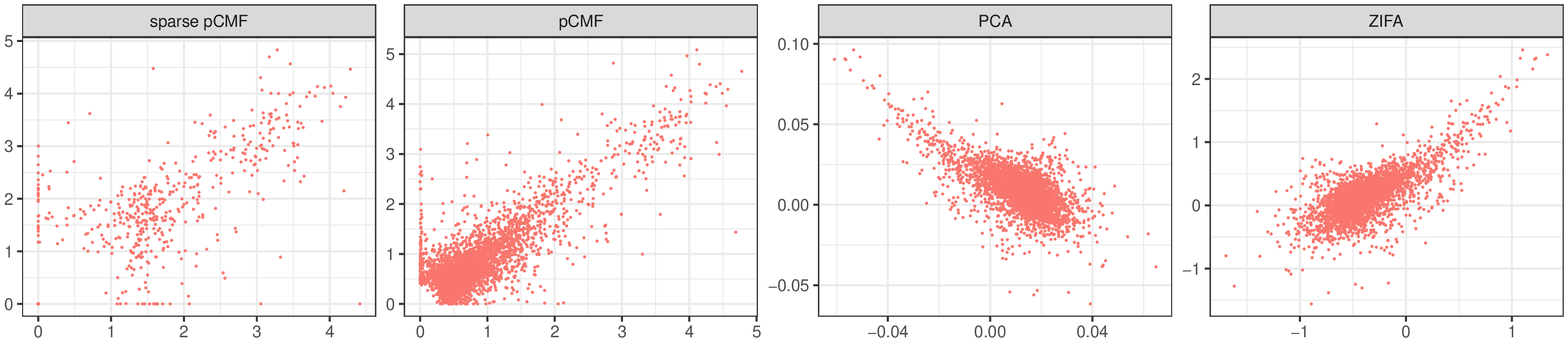}
\caption{Analysis of the goldstandard scRNA-seq data from \cite{freytag2018}, $925$ cells, $8580$ genes. Visualization of the genes in a latent space of dimension 2.}
\label{supp:fig:data:visu:gold:genes}
\end{figure}

\begin{figure}[h]
\centering
\includegraphics[width=.99\linewidth]{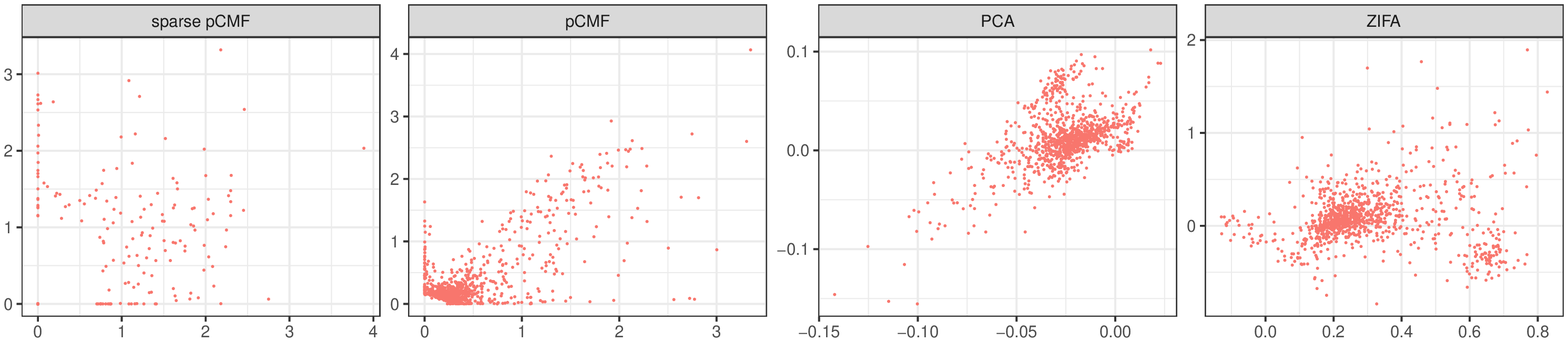}
\caption{Analysis of the silverstandard 5 scRNA-seq dataset from \cite{freytag2018}, $8352$ cells, $4547$ genes. Visualization of the genes in a latent space of dimension 2.}
\label{supp:fig:data:visu:silver:genes}
\end{figure}

\begin{figure}[h]
\centering
\includegraphics[width=.99\linewidth]{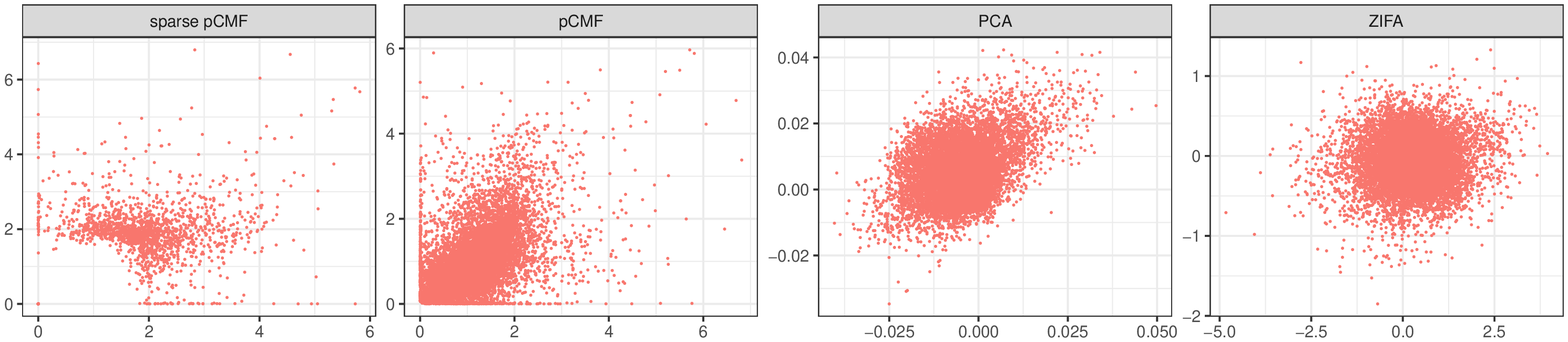}
\caption{Analysis of the scRNA-seq dataset from \cite{llorens-bobadilla2015}, $141$ cells, $13826$ genes. Visualization of the genes in a latent space of dimension 2.}
\label{supp:fig:data:visu:llorens:genes}
\end{figure}

See \cref{supp:fig:data:visu:baron:genes,supp:fig:data:visu:gold:genes,supp:fig:data:visu:silver:genes,supp:fig:data:visu:llorens:genes}.

\end{document}